\documentclass{aa}  
\usepackage{graphicx}
\usepackage{xcolor}
\usepackage{hyperref} 
\usepackage{multirow} 
\usepackage{booktabs} 
\usepackage{longtable} 

\newcommand{\teff}{\ensuremath{T_{\mathrm{eff}}}\xspace}
\newcommand{\kms}{\ensuremath{\rm{km}\,s^{-1}}\xspace}
\newcommand{\logg}{\ensuremath{\log g}\xspace}
\newcommand{\feh}{\rm{[Fe/H]}\xspace}

\usepackage{txfonts}
\hypersetup{final} 
\begin{document} 

   \title{The X-Shooter Spectral Library (XSL): Data Release 2
   \thanks{Table~\ref{tab:sample} and the spectra are available in electronic form
at the CDS via anonymous ftp to \url{cdsarc.u-strasbg.fr (130.79.128.5)} or via \url{http://cdsweb.u-strasbg.fr/cgi-bin/qcat?J/A+A/2019/36825}}
 \fnmsep
   \thanks{Based on observations collected at ESO Paranal La Silla Observatory, Chile, Prog. IDs 084.B-0869, 085.B-0751, and 189.B-0925 (PI Trager).}}

  \author{Ana\"{i}s Gonneau\inst{\ref{inst1},\ref{inst2},\ref{inst4}}
            \and M. Lyubenova\inst{\ref{inst3},\ref{inst2}}
            \and  A. Lan\c{c}on\inst{\ref{inst4}}
            \and S.~C. Trager\inst{\ref{inst2}}
            \and R.~F. Peletier\inst{\ref{inst2}}
            \and \\ A. Arentsen\inst{\ref{inst7},\ref{inst2}}
            \and Y.-P. Chen\inst{\ref{inst6}}
            \and P.~R.~T. Coelho\inst{\ref{inst11}}
            \and M. Dries\inst{\ref{inst2}}
            \and J. Falc{\'o}n-Barroso\inst{\ref{inst8},\ref{inst9}}
            \and \\P. Prugniel\inst{\ref{inst5}}
            \and P. S{\'a}nchez-Bl{\'a}zquez\inst{\ref{inst10}}
            \and A. Vazdekis\inst{\ref{inst8},\ref{inst9}}
            \and K. Verro\inst{\ref{inst2}}
          }

\institute{Institute of Astronomy, University of Cambridge, Madingley Road, Cambridge CB3 0HA, United Kingdom \\
    \email{agonneau@ast.cam.ac.uk} \label{inst1}
    \and
	Kapteyn Astronomical Institute, University of Groningen, Landleven 12, 9747 AD Groningen, the Netherlands \label{inst2}
	\and
    ESO, Karl-Schwarzschild-Str. 2, D-85748 Garching bei  M\"unchen, Germany \label{inst3}
    \and
    Observatoire Astronomique de Strasbourg, Universit\'e de Strasbourg, CNRS, UMR 7550, 11 rue de l'Universit\'e, \\F-67000 Strasbourg, France\label{inst4}
    \and
    CRAL-Observatoire de Lyon, Universit\'e de Lyon, Lyon I,  CNRS, UMR5574, France \label{inst5}
    \and
    New York University Abu Dhabi, Abu Dhabi, P.O. Box 129188, Abu Dhabi, United Arab Emirates\label{inst6} 
    \and 
    Leibniz-Institut f\"{u}r Astrophysik Potsdam (AIP), An der Sternwarte 16, D-14482 Potsdam, Germany \label{inst7}
    \and 
    Instituto de Astrof\'isica de Canarias, V\'ia L\'actea s/n, La Laguna, Tenerife, Spain\label{inst8}
    \and
    Departamento de Astrof\'isica, Universidad de La Laguna, E-38205 La Laguna, Tenerife, Spain\label{inst9}
    \and
    Departamento de F\'isica de la Tierra y Astrof\'isica, UCM, 28040 Madrid, Spain\label{inst10}
    \and
    Universidade de S\~{a}o Paulo, Instituto de Astronomia, Geof\'isica e Ci\^{e}ncias Atmosf\'ericas, Rua do Mat\~{a}o 1226, 05508-090, S\~{a}o Paulo, Brazil\label{inst11}
    }

   \date{Received 2 October 2019; Accepted 29 December 2019}

\abstract{We present the second data release (DR2) of the X-Shooter Spectral Library (XSL), which contains all the spectra obtained over the six semesters of that program. This release supersedes our first data release from \citeyear{Chen14}, with a larger number of spectra
(813 observations of 666 stars) and with a more extended wavelength coverage as the data from the near-infrared arm of the X-Shooter spectrograph are now included. The DR2 spectra then consist of three segments that were observed simultaneously and, if combined, cover the range between $\sim$300\,nm and $\sim$2.45\,$\mu$m at a spectral resolving power close to $R=10\,000$. The spectra were corrected for instrument transmission and telluric absorption, and they were also corrected for wavelength-dependent flux-losses in 85\% of the cases. 
On average, synthesized broad-band colors agree
with those of the MILES library and of the combined IRTF and Extended IRTF libraries to within $\sim\!1$\%.
The scatter in these comparisons indicates typical errors on individual colors in the XSL of 2$-$4\,\%. 
The comparison with 2MASS point source photometry shows systematics of up to 5\,\% in some colors, which we attribute mostly to zero-point or transmission curve errors and a scatter that is consistent with the above uncertainty estimates.
The final spectra were corrected for radial velocity and are provided in the rest-frame (with wavelengths in air). The spectra cover a large range of spectral types and chemical compositions (with an emphasis on the red giant branch), which makes this library an asset when creating stellar population synthesis models or for the validation of near-ultraviolet to near-infrared theoretical stellar spectra across the Hertzsprung-Russell diagram. }


   \maketitle
%


\section{Introduction}
\label{sec:intro}
 
Stellar spectral libraries are fundamental resources that shape our understanding of stellar astrophysics and allow us to study the stellar populations of galaxies across the Universe. These libraries come in two flavors: empirical, which are composed of a well-defined set of stars with certain stellar atmospheric parameters cove\-rage, and theoretical where stellar spectra are computed for an arbitrarily large set of parameters and extensive wavelength coverage. The list of empirical and theoretical spectral libraries grows continuously due to their versatile usage in modern astrophysics and they are boosted by developments of new instrumentation\footnote{For an extensive list of stellar spectral libraries see David Montes' web collection at \url{https://webs.ucm.es/info/Astrof/invest/actividad/spectra.html} and references therein.}(e.g., \citealt{Coelho09, Husser2013, Allende18} for the theoretical side and \citealt{Vazdekis16, Villaume17} for the empirical side).

For the purpose of building stellar population synthesis mo\-dels, 
a good stellar library should have four key properties, as highlighted by \citet{trager12}: an exhaustive coverage of all stellar evolutionary phases and chemical compositions to represent as well as possible the integrated light of real stellar systems; a broad wavelength coverage since not all stellar phases contribute equally at all wavelengths; simultaneous observations at all wavelengths to avoid issues due to temporal stellar spectral variations; and good calibration of the individual stars in terms of  flux and wavelength calibration as well as  high-precision stellar atmospheric parameters.
This latter point requires comparison with synthetic stellar spectra, which highlights another important application of libraries with the above properties: the va\-lidation and improvement of theoretical stellar models across the Hertzsprung-Russell (HR) diagram and across wavelengths.

These were  the goals of the X-Shooter Spectral Library (XSL), which is a moderate-resolution ($R\sim10\,000$) spectral library that was designed to cover most of the HR diagram. The observations were carried out with the X-Shooter three-arm spectrograph on ESO's VLT \citep{Vernet11} in two phases, a pilot and an ESO Large Program, spanning six semesters in total. In our first data release \citep[][hereafter DR1]{Chen14}, we present spectra of 237 unique stars, which were observed during the pilot program, for a wavelength range that was restricted to the two optical arms of X-Shooter 
(300--1024 nm).

In the current data release (DR2), we present our full set of 813 observations of 666 stars, now also including data from the near-infrared arm of X-Shooter. This near-infrared extension is undoubtedly one of the main advantages of XSL over other empirical spectral libraries in the literature. Empirical near-infrared libraries for a relatively wide range of stellar parameters have been constructed in the past, first at very low spectral resolution \citep[e.g.,][]{JM70, LRV1992}, then progressively at intermediate resolution \citep[e.g.,][]{LW2000, Ivanov04,  Rayner09,Villaume17}, or at higher spectral resolution but in restricted wavelength ranges \citep[e.g.,][]{Cenarro01, MajewskiAPOGEE}. We note that some of these libraries did not attempt to preserve the shapes of the stellar continua across the wavelengths observed. Only some have been combined with optical libraries for the purpose of calculating the spectra of synthetic stellar populations \citep[e.g.,][]{Pickles98, lancon99, vaz2003, maraston05, LanconM82_2008, RoeckVaz2016, ConroyVillaume2018}. In these efforts, the need to merge optical and near-infrared observations of distinct samples of stars is an inevitable cause of systematic errors. The X-Shooter instrument helps alleviate this issue, as it acquires simultaneous observations of the whole spectral range for every stellar target. In addition, the spectral resolution achieved is higher than those of any of the previous libraries that cover as broad  spectral range.

The XSL DR2 data are homogeneously reduced and cali\-brated, and the spectra are made available in three spectral ranges, corresponding to the three arms of the spectrograph: UVB - 300--556 nm, VIS - 533--1020 nm, and NIR - 994--2480 nm. Our data release contains, similar to DR1, repeated observations of several cool giants stars. Both data releases are available on our website: \url{http://xsl.astro.unistra.fr}.

Our sample selection and observing strategy are described in Section~\ref{sec:sample}. Section~\ref{sec:data_red} gives details about the data reduction and calibration process. The final spectra are assessed in terms of spectral resolution and energy distribution in Section~\ref{sec:qa}. In Section~\ref{sec:data_prod}, we summarize the data products made available, before concluding in Section ~\ref{sec:concl}. Additional details about the input catalogs used to build this library can be found in Appendix~\ref{app:input_cats}.
Appendix~\ref{app:comments_stars} collects peculiarities of some of the stars in the program, and observational artifacts that affect some of the spectra. 
Finally, Appendix~\ref{app:list_stars} provides the log of all the observations.


\section{Sample selection and observing strategy}
\label{sec:sample}

\subsection{Selection criteria}

\begin{figure}[!h]
    \centering
    \includegraphics[width=\linewidth,trim=0cm 3cm 0cm 3cm,clip]{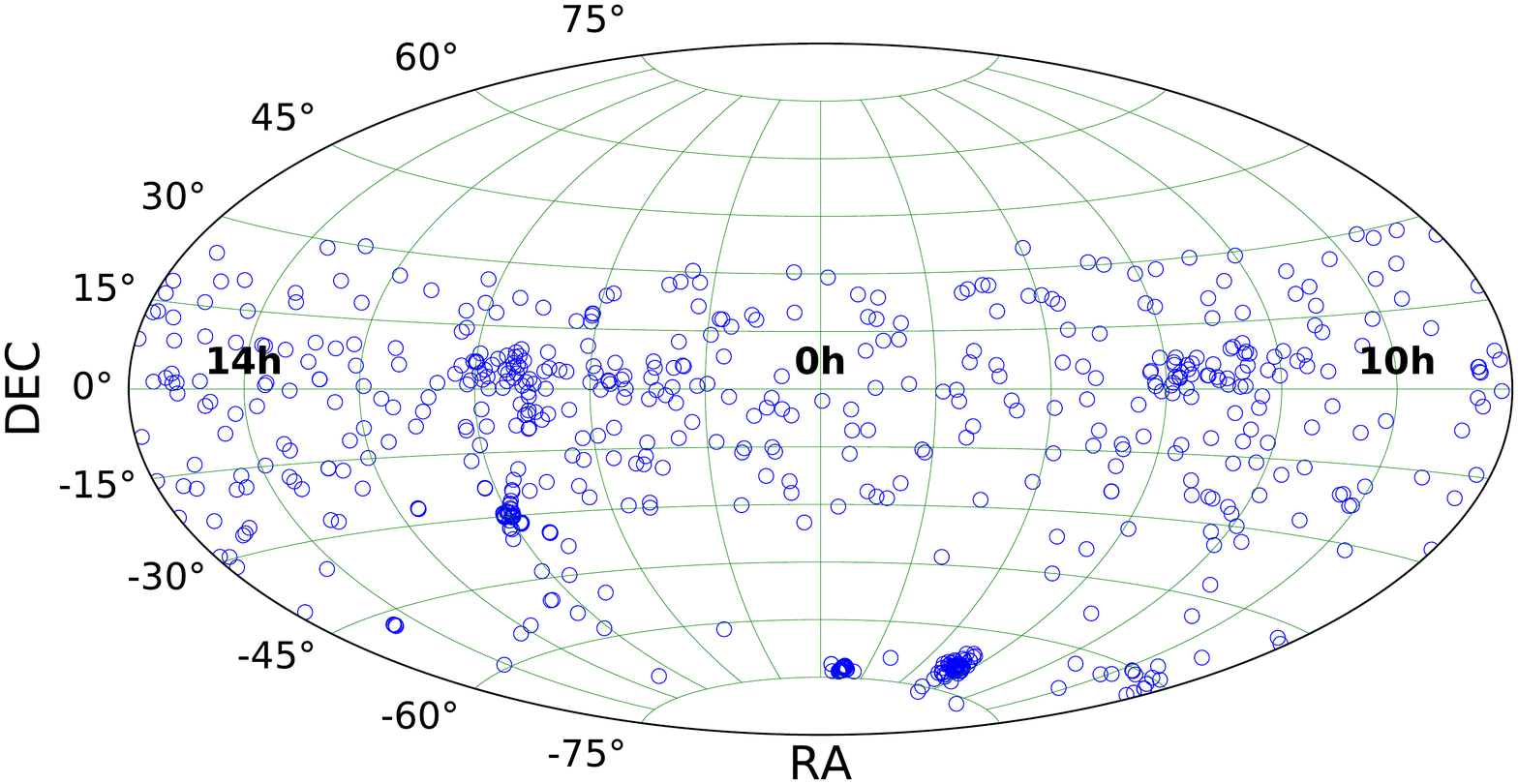}
    \caption{Positions of XSL stars in the sky (Aitoff projection). }
    \label{fig:ra_dec}
\end{figure}

The XSL target stars were selected to cover as much of the Hertzsprung–Russell diagram as possible in the allocated time, with a wide range of chemical compositions. Our original sample consists of 679 unique stars. The primary references for the construction of the XSL target list were existing spectral libraries or compilations of stellar parameter measurements. In particular, the XSL sample has a strong overlap with the MILES spectral library \citep[142 stars,][]{Sanchez06,Cenarro07} and the NGSL library \citep[135 stars, ][]{Gregg06}. We completed the list with objects from the parameter compilation PASTEL \citep{Soubiran10, Soubiran16}
and from a variety of more specialised catalogs (see Table \ref{tab:inputcats} in the Appendix for details and corresponding references).
In Figure~\ref{fig:ra_dec} we plot the distribution of the selected XSL stars on the sky.

More than half of the stars in the XSL library are red giants in the broad sense, which includes red supergiants or asymptotic giant branch stars. These stars provide strong (age-dependent) contributions to the near-infrared emission of galaxies \citep[e.g.,][]{lancon99,maraston05,melbourne12}. The luminous red stars in XSL are located in star clusters, in the field, in the Galactic bulge and in the Magellanic Clouds (Table\,\ref{tab:inputcats}). 
A fraction of the asymptotic giant branch stars are carbon stars, whose spectra have been studied in detail by \cite{Gonneau16,Gonneau17}.
We obtained repeated observations for a fraction of the XSL stars ($\sim$ 20\% of our sample), mainly in the cases of luminous cool stars, known or suspected to vary in time.

\begin{figure}[!ht]
 \centering
\includegraphics[width=\linewidth,trim=0.5cm 0.5cm 1.5cm 0.2cm,clip]{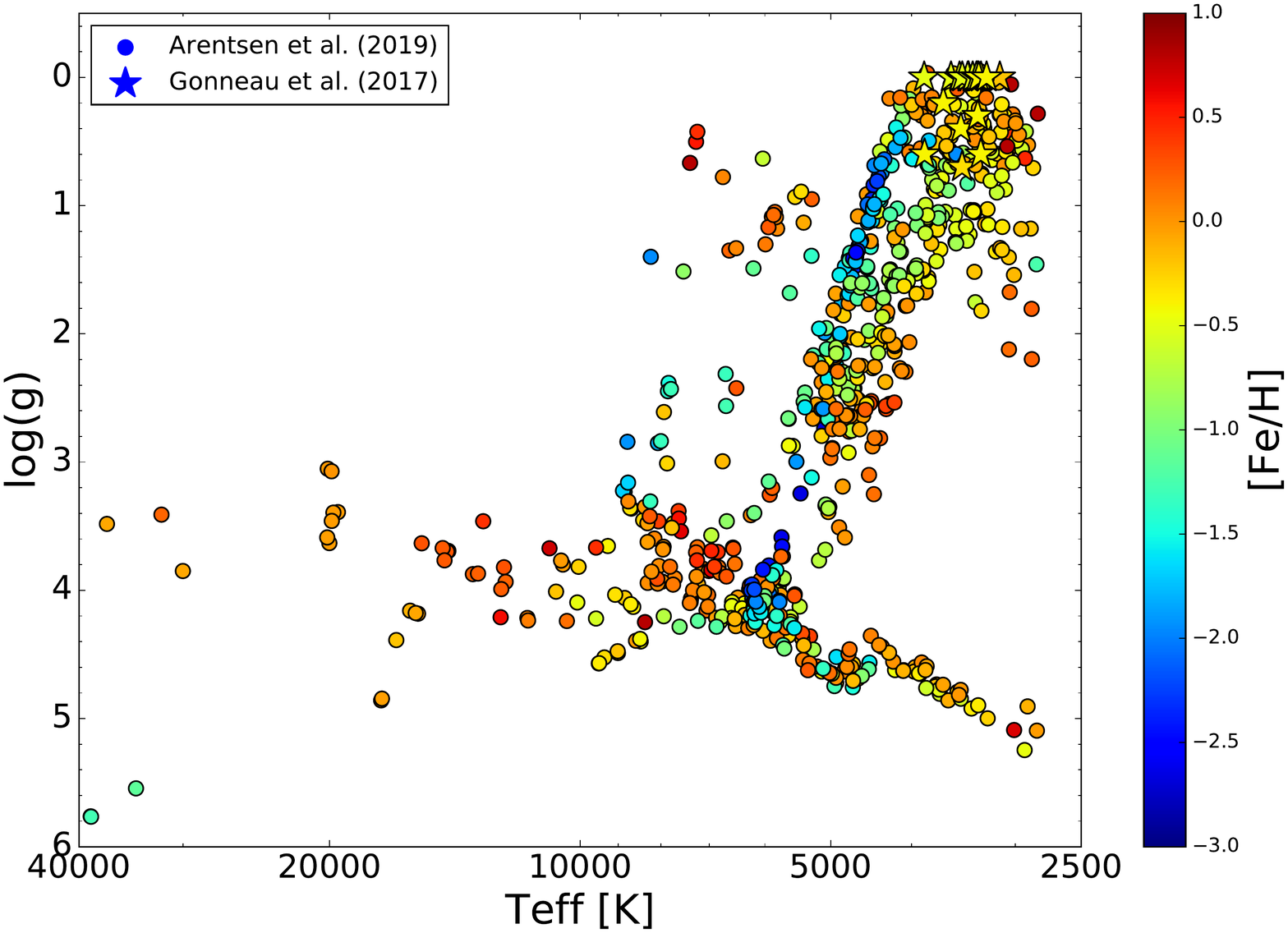}
  \caption[Stars from XSL]{Stellar atmospheric parameters, calculated by \citet[][open symbols]{Arentsen19} and \citet[][star symbols]{Gonneau17}, for 769 XSL spectra. }
  \label{fig:xsl_hr}
\end{figure}

Figure~\ref{fig:xsl_hr} presents the stellar atmospheric parameters cove\-rage of the XSL sample as determined by \citet{Arentsen19} and \citet{Gonneau17}. We refer the reader to Figure~2 of \citet{Chen14} where the HR diagram is presented using lite\-rature values. \citet{Arentsen19} estimated atmospheric parameters in a homogeneous manner for 754 XSL spectra of 616 stars: they fit the ultraviolet and visible spectra with the ULySS package \citep{Koleva2009} using the MILES spectral interpolator \citep{Sharma16} as a reference.
\citet{Gonneau17} determined atmospheric  parameters of carbon-rich stars
by comparing the observations to a set of high-resolution synthetic carbon star spectra, based on hydrostatic model atmospheres \citep{aringer16}\footnote{Only the stars less affected by pulsation \cite[namely stars from Groups A, B and C as defined in][]{Gonneau17} are plotted in Fig.~\ref{fig:xsl_hr}.} .

The aim of the selection was to obtain representative spectra mapping the widest area of the parameter space defined by the effective temperature (\teff), the surface gravity (\logg), and the iron metallicity (\feh). Therefore, the sample does not reproduce the natural frequencies of various stellar types.
At the time the target lists were prepared, large public catalogs of homogeneous stellar parameter estimates were still lacking, and we were unable to populate certain regions of parameters in a satisfactory way. In particular, some of the Bulge stars selected on the basis of high estimated metallicities turned out
 to be less metal-rich than expected \citep{Arentsen19}. 

For applications requiring a higher level of completeness, we note that the X-Shooter instrument archive has become a useful source of complementary data over the years. Low mass dwarfs from the archive were combined with XSL data for population synthesis applications by \citet{Dries19}. The red supergiants of \citet{Davies13} were observed with settings similar to XSL and can usefully complement the XSL library in that region. Finally, the ESO archive\footnote{\url{http://archive.eso.org/wdb/wdb/eso/xshooter/form}} contains a very large number of observations of B-type stars, because these have been used over many years for the correction of telluric absorption. We do not discuss these external data in this article.

\subsection{Observing strategy}

We gathered the XSL observations in two phases: a pilot program (periods P84 and P85, October 2009 -- September 2010) and a Large Program over four semesters (P89-P92, April 2012 -- March 2014). 

We observed each science target with two slit widths. We first took a spectrum with a narrow slit to achieve the required spectral resolution. The narrow-slit widths for the UVB, VIS and NIR arms of X-Shooter were 0.5\arcsec, 0.7\arcsec, 0.6\arcsec, respectively, leading to nominal resolutions $R\!=\!\lambda/\Delta\lambda \sim9200, \sim11000$ and $\sim7770$. 
With the narrow slits, a wavelength-dependent fraction of the stellar flux is lost. The loss-fraction depends on the seeing, but also on the centering of the star 
in the slit and on the performance of the atmospheric dispersion corrector of the instrument, both of which are not controlled precisely. 
The most reliable approach to correct these losses is to observe the targets with a wide slit positioned vertically on the sky. 
That observation can be calibrated into absolute flux when conditions are photometric and then can be used to correct the energy distribution of the narrow-slit observation of the same target. In non-photometric conditions, as tolerated in our observations (``thin cirrus'' tolerance of the ESO VLT scheduling), absolute fluxes cannot be garanteed, but the shape of the energy distribution is properly recovered with this method under the assumption that the cirrus have a flat transmission curve\footnote{Using the extinction curve for normal cirrus of Lynch \& Mazuk (2001), we find that the ratio of V-band to K-band transmission varies by less than 1\,\% when transmission drops from 100\,\% to 60\,\% because of standard cirrus.}. We took wide-slit spectra for every target immediately after the narrow-slit ones, using a 5\arcsec wide slit.

\subsubsection{Observing mode}

We performed our observations using the SLIT spectroscopy mode of X-Shooter. In this mode three observing strategies are available: STARE, NODDING, and OFFSET. We used each of them depending on the desired outcome.
In STARE mode the star was located at the center of the slit and the sky background was estimated from each side of the stellar signal on the observed frame. We used this mode for the wide-slit observations.

The NODDING acquisition mode allows observations of the star at two positions (A and B) along the spectrograph slit. In this way we achieved an improved sky subtraction with a double-pass subtraction.
Almost all our narrow-slit observations were taken in NODDING mode. 
An OFFSET acquisition mode alternates between the star and an empty sky region. We used this mode for 12 observations of very bright stars in the pilot program.

\subsection{Spectro-photometric standard stars}

In order to flux-calibrate our science spectra we used spectro-photometric standard stars (typically white dwarf stars of type DA). These stars were observed once per night using a 5\arcsec slit width. 

In practice, we used only the five following stars, for which the spectral energy distribution and other parameters are given in the X-Shooter manual\footnote{\url{http://www.eso.org/sci/facilities/paranal/instruments/xshooter/tools/specphot_list.html}}: BD+17\,4708, EG\,274, Feige\,110, LTT\,3218, and LTT\,7987.
The other reference stars (CD-32 9927, EG 21, GD 153, GD 71, LTT 1020 and LTT 4364) were discarded either because their spectral energy distribution was not reliably known or because the long exposure times of their default observations (300 or 600 se\-conds) resulted in saturated NIR $K$-band spectra.

\section{Data reduction and calibration}
\label{sec:data_red}

We have reduced the full set of XSL observations over the pilot and the Large Programs in a uniform manner. This allows us to provide an updated calibration of the spectra initially released under DR1 and to ensure a homogeneity of the present DR2.

\begin{figure}[t]
 \centering
 \fbox{\includegraphics[width=0.96\linewidth,clip]{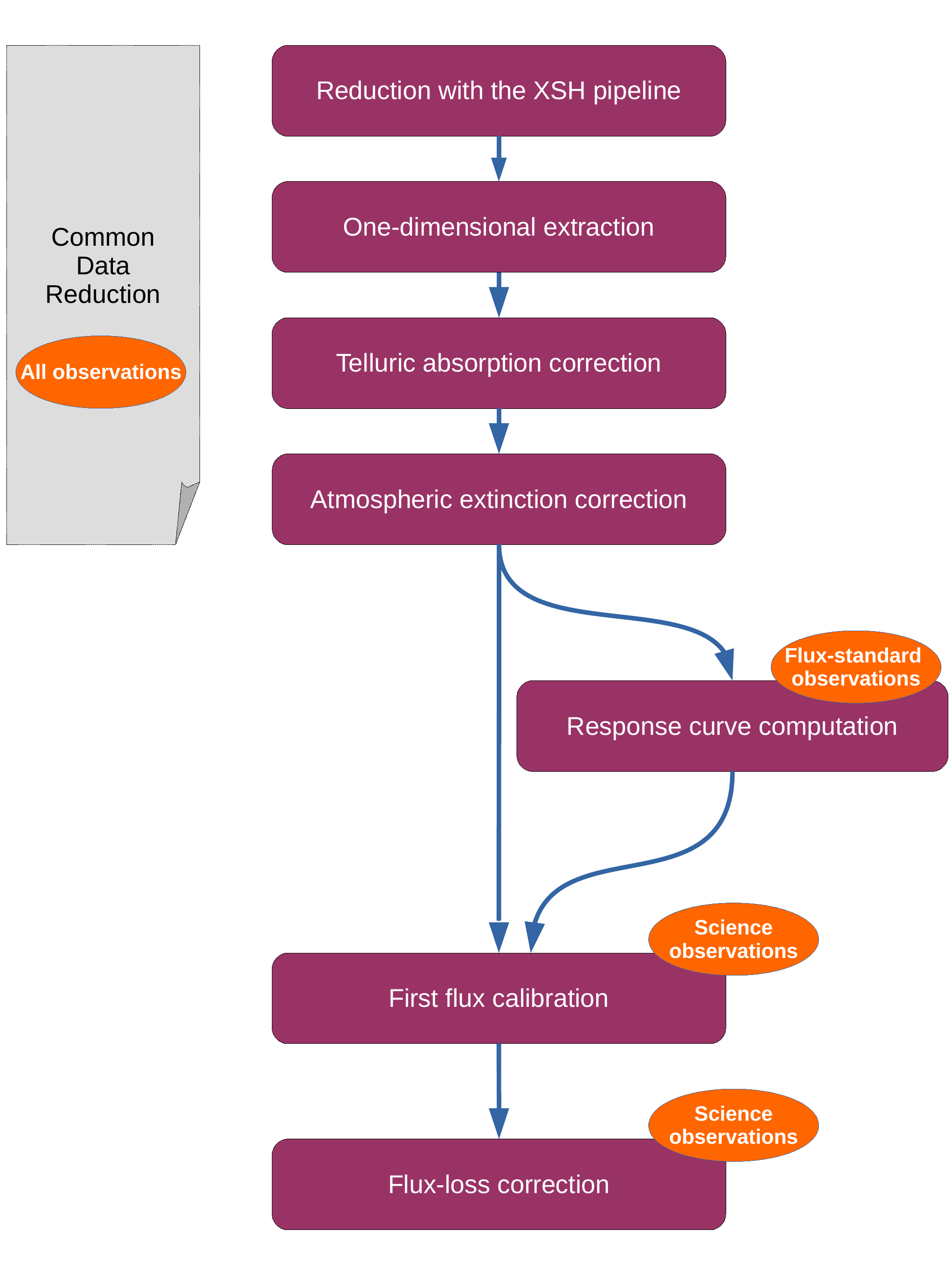}}  
  \caption{Overview of the XSL data reduction process. }
  \label{fig:xsl_red}
\end{figure}

Our data reduction and calibration processes are illustrated in Fig.~\ref{fig:xsl_red}. We apply the first four steps (referred to as ``common data reduction'') to all observations: science target stars and flux standards, wide and narrow slits, and all spectral arms. 
We use the X-Shooter data reduction pipeline \citep[hereafter XSH pipeline, Sect.~\ref{sec:xsh_pipeline}]{Modigliani10}, 
up to the creation of two-dimensional (2D) order spectra.
 Then we extract  1-dimensional (1D) spectra outside of the XSH pipeline to better control the rejection of  bad pixels (see Sect.~\ref{sec:extract1D}). 
We further apply corrections for the telluric absorption bands (Sect.~\ref{sec:atmos_molec}) and for the continuous atmospheric extinction (Sect.~\ref{sec:atmos_scatter}).

After processing all observations in this uniform way, we derive response curves from the flux-standard spectra (Sect.~\ref{sec:response_curve}), and we use these to process both the narrow and wide slit science spectra
(Sect.~\ref{sec:response_curve}).
Finally, using the wide-slit spectrum corresponding to each narrow slit observation, we derive a wavelength dependent flux-loss correction and apply it to achieve the final XSL spectrum for each stellar observation (Sect.~\ref{sec:flux_loss}).

\subsection{XSH Pipeline}
\label{sec:xsh_pipeline}

The XSH pipeline consists of a set of data processing modules that perform the individual tasks of the data reduction. These  ``recipes'' can be chained in a flexible way to fulfill the requirement of a specific program. We use the XSH pipeline version 2.6.8, in \textit{physical} mode as per the pipeline recommendation. In that mode, the wavelength and spatial scale calibrations are performed by optimising a physical model of the instrument.
To achieve a more flexible automation of our process we run the XSH pipeline via the EsoReflex environment\footnote{\url{https://www.eso.org/sci/software/esoreflex/}} (version 3.12).


For each science observation, the input to the XSH pipeline is a set of raw data-frames that we identify and collect using the associations of calibration files provided by the ESO archive.
When necessary, we modify the default associations to ensure that the flux-standard frames are always reduced with the same flat-field as the corresponding science frames, rather than with the flat-field nearest-in-time (see Sect.~\ref{sec:response_curve}). 
Satu\-ration was a known problem for several observations of the pilot program but occurred rarely in later semesters. Here, we apply a scheme similar to that of DR1 to flag strongly saturated observations automatically and exclude these from further processing.


We used the default parameters for the creation of the calibration frames (up to the \textit{xsh\_flexcomp} recipe).
To transform the science and flux-standard frames into flat-fielded, rectified and wavelength-calibrated 2D order spectra, we used recipes
 \textit{xsh\_scired\_slit\_stare} for wide-slit observations, and 
\textit{xsh\_scired\_slit\_offset} for narrow-slit observations.
In particular, we reduced any frames observed in NODDING mode as if they were taken in OFFSET mode, which gives us more flexibility in the handling of bad pixels and improves the quality of our final spectra. In STARE mode, we switch off the sky-subtraction offered by the XSH pipeline because we apply our own algorithm during the subsequent 1D-extraction step.
Two key para\-meters of the above recipes control the sampling of the output spectra
along the wavelength axis and along the spatial axis.
We set them respectively to 0.015 nm and 0.16\arcsec\ for the UVB and VIS arms, and to 0.06 nm and 0.21\arcsec\ for the NIR arm.  

%
%
The final products of the reduction with the X-Shooter pipeline are 2D frames containing the spectral orders of each arm. They include a map of pixel quality flags and a map of the estimated variances of the pixel errors, which we use to propagate errors through subsequent processing steps.

\subsection{1D extraction}
\label{sec:extract1D}

A main driver for our choice of performing the extraction of 1D spectra outside of the XSH pipeline was the need for more control over the rejection of bad pixels and the sky subtraction. Our extraction procedure follows a prescription adapted from \citet{Horne86}. It implements a rejection of masked and outlier pixels, as well as a weighting scheme based on a smooth throughput profile, that can follow residual distortions in the rectified 2D-spectra, and on the local pixel-variance as estimated by the earlier steps in the ESO pipeline. 
The extraction procedure is an updated version of the one presented in \citet{Gonneau16}, and the description below focuses on the modified elements. 

\subsubsection{Fraction of bad pixels}

Before extracting a given order, we double-checked for saturation and other invalid pixels by counting the percentage of pixels flaged as bad along the spectrum. The count is restricted to a box extending $\pm$3\,/\,$\pm$2 pixels (respectively for the UVB/VIS arms and the NIR arm) to either side of the peak of the spatial profile of the spectrum. As a general rule, the extraction process is abandoned if more than 30\% of the pixels are determined to be bad. 
 Exceptions are implemented for pixels with XSH pipeline-code 6 (unremoved cosmic ray), a warning that we discard to improve the extraction of the observations taken in exceptionally good seeing (see Section\,\ref{sec:chess}), and for XSH pipeline-code 21 (extrapolated flux in the NIR) in the longest wavelength order of the NIR arm, because keeping those pixels was found to reduce discontinuities between the two last orders of that arm significantly \citep[this is an update with respect to][who describe the discontinuities]{Gonneau16}.

\subsubsection{Sky background removal}

We carried out a sky subtraction for all spectra acquired in STARE mode, as well as for any long-exposure VIS and NIR frames observed in NODDING and OFFSET modes (science frames with exposure times $\geq$ 600\,s in the VIS and exposure time $\geq$ 200\,s in the NIR; flux standard frames with exposure time $\geq$ 190\,s in the VIS and all the NIR frames).
The areas along the slit from which the median sky levels are computed are selected in each frame after fitting a Gaussian to the spatial profile of the star, in order to account for the seeing. In STARE mode, these adjustments avoid artifacts such as the negative fluxes that sometimes result from over-subtraction in the standard XSH-pipeline process. In long-exposure spectra, the main purpose is to account for variations in the sky background between the two combined acquisitions.

\subsubsection{Updated standard extraction}

\citeauthor{Horne86}'s extraction scheme relies on obtaining first a rough guess for the 1D spectrum (that we refer to as the \textit{standard} spectrum) and then optimi\-zing this extraction with an inverse-variance weighted algorithm, that accounts for the spatial profile of the light at a given wavelength (we call the result the \textit{final} spectrum).

To obtain the standard spectrum, we summed the (sky-subtracted) 2D spectrum along the spatial dimension over a limited range of pixels. We updated our procedure by setting the limits to $\pm$ 4 or 8 pixels, respectively for NODDING and STARE/OFFSET frames, centered on the peak of the signal. 
The calculation of the final spectrum is as in \citet{Gonneau16}. Along the spatial axis, the ratio of the flux in a pixel to the normalized spatial profile is taken as an estimate of the stellar flux at the wavelength of interest and these estimates are averaged with an inverse-variance weight. A limited number of iterations are implemented to add outliers to the bad-pixel mask (maximum 3 at any wavelengths), and these pixels are discarded in the final sum.

\subsubsection{Merging of the individual orders}

\begin{figure}[!h]
  	\centering
    \includegraphics[width=.95\linewidth,clip]{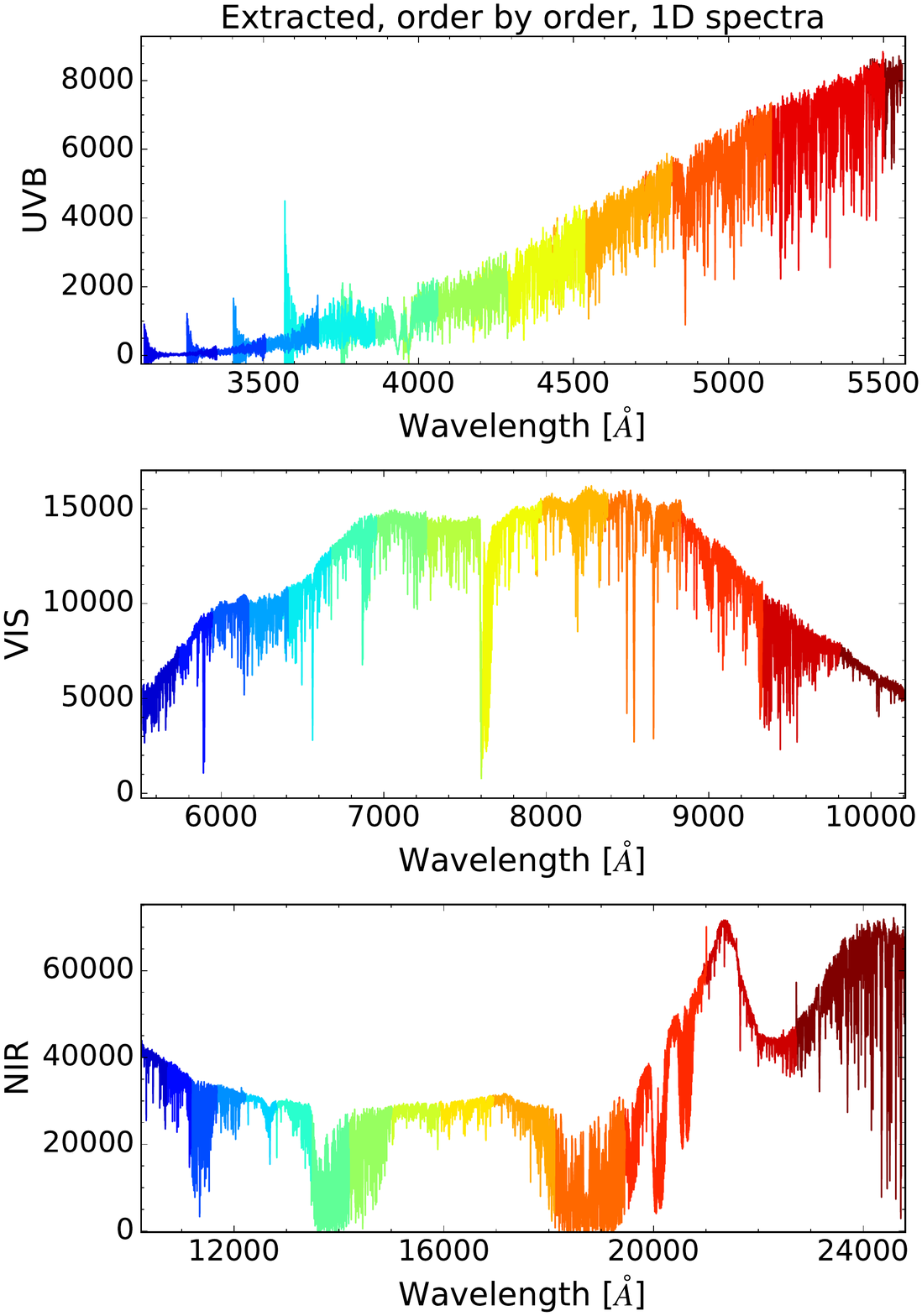}
    \caption{Examples of order-by-order spectra for the three arms (XSL observation X0365). From top to bottom: UVB (12 orders), VIS (15 orders) and NIR (16 orders).}
\label{fig:order_by_order}
\end{figure}

Figure~\ref{fig:order_by_order} shows the overlapping orders before mer\-ging for each arm of the spectrograph. The flat-fielding algorithm of the XSH pipeline normalizes flat-field frames only globally, hence the blaze function of the instrument is removed in the division by the flat-field, and the extracted orders align well.
To merge them, we combined the spectra in the regions of overlap, using weights that combine the inverse variance and a coefficient, $\alpha$,
set to vary linearly from 0 to 1 across those regions. 
This choice was made because we found that the single-order variance was frequently underestimated at the very end of orders. Hence the flux of each overlapping order, $F_{merging}$, is calculated as follows: 
\begin{equation}
F_{merging} = \frac{ (1-\alpha) * F_L/V_L  + \alpha * F_R/V_R}{((1-\alpha)/V_L + \alpha/V_R)}
\end{equation},
where $F$ and $V$ stand for pixel fluxes and variances, and subscripts $L$ and $R$ refer to the spectral orders on the left and right sides of region of overlap.

For the frames observed in NODDING mode, the 1D extraction procedure produces two separate spectra that we ave\-raged together. 
The end product of this procedure is a continuous 1D spectrum with two extensions: a flux spectrum and an error spectrum.


\subsubsection{Chessboard effect}
\label{sec:chess}

\begin{figure}[!h]
 \centering
    \includegraphics[width=.95\linewidth,trim=0 0 0 0,clip]{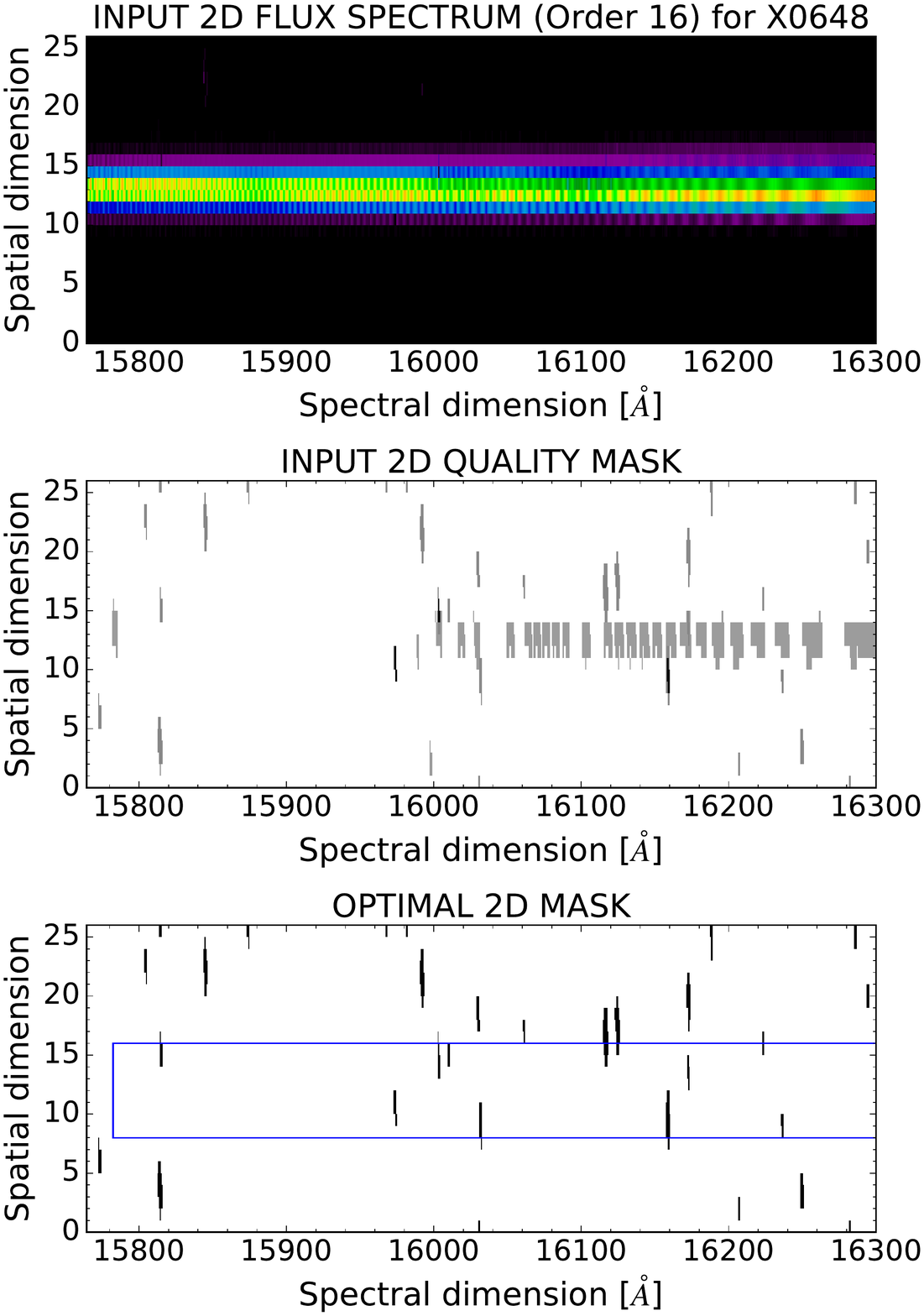}
  \caption{Example of ORDER2D frame affected by the chessboard effect. \textbf{Top:} Input 2D flux spectrum, before extraction. 
  \textbf{Middle:} Input 2D qua\-lity mask. \textbf{Bottom:} Optimal 2D mask.
  For both 2D masks, the bad pixels are shown in black  and gray.}
  \label{fig:chess_raw}
\end{figure}

Some of our NIR frames are affected by an issue that we call the ``chessboard effect''. 
When the seeing is exceptional (typically  $<$ 2 pixels per FWHM of the spectral profile in parts of the raw data), the interpolation scheme used by the XSH pipeline for the geometric transformation into a rectified, wavelength calibrated frame does not perform well. The resulting images (known as ORDER2D frames in the pipeline) display a chessboard pattern as illustrated in the top panel of Fig.~\ref{fig:chess_raw}. Because the wavy patterns seen along single lines of the rectified 2D-spectrum vary abruptly between one line and the next, the crests of these waves are mistaken for cosmic ray hits. 

\begin{figure}[!h]
  	\centering
   \includegraphics[width=0.95\linewidth, clip]{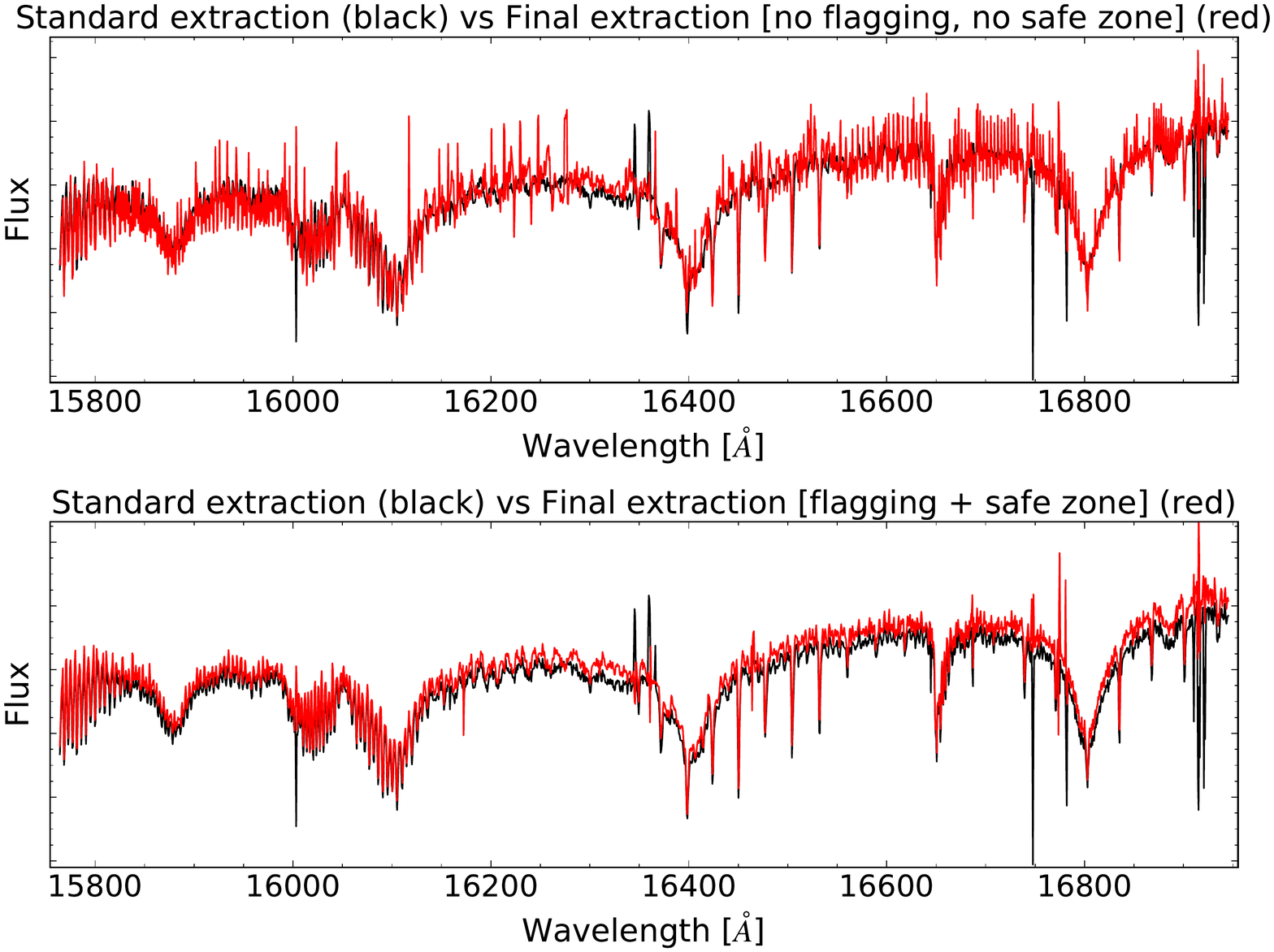}
   	\caption{Examples of extracted spectra with the chessboard effect (X0648).
    The red spectra correspond to the inverse-variance weighted extractions, while the black spectra are for the standard extractions. The original extraction is shown on the upper panel, and the updated extraction to deal with the chessboard effect is shown on the lower panel (flagging and safe zone in the rejection scheme).  } 
\label{fig:chess_ext}
\end{figure}

To deal with this, we first identify pixels with bit code 6 in the quality-mask images produced by the XSH pipeline (flag 6 = unremoved cosmic rays, cf. the bad pixel code conventions in the X-Shooter Pipeline Manual) and reset these flags to ``good pixel'' \footnote{A similar solution is now recommended by ESO on the X-Shooter FAQ webpage: \scriptsize \url{http://www.eso.org/sci/data-processing/faq/my-spectrum-of-a-bright-source-looks-much-noisier-than-i-expected-why.html}} \normalsize. Then, to avoid any re-flagging by our own  extraction procedure, we define a ``safe zone'' in the 2D frame in which any cosmic-ray identification is switched off. This rectangular zone extends to $\pm$ 4 or 6 pixels on either side of the peak of the spatial profile of the spectrum (respectively for NODDING or for  STARE/OFFSET frames). This scheme was also used in a few VIS frames. The bottom panels of Fig.~\ref{fig:chess_raw} show an example of a 2D quality-mask before and after editing, and the extent of the safe zone (represented as a blue rectangle). 
Figure~\ref{fig:chess_ext} shows a comparison of the different extraction methods, with and without the flagging and the safe zone, for the ORDER 16 of the NIR arm.

\subsection{Telluric absorption correction}
\label{sec:atmos_molec}

\begin{figure*}[!ht]
	\centering
	{\includegraphics[width=0.9\textwidth]{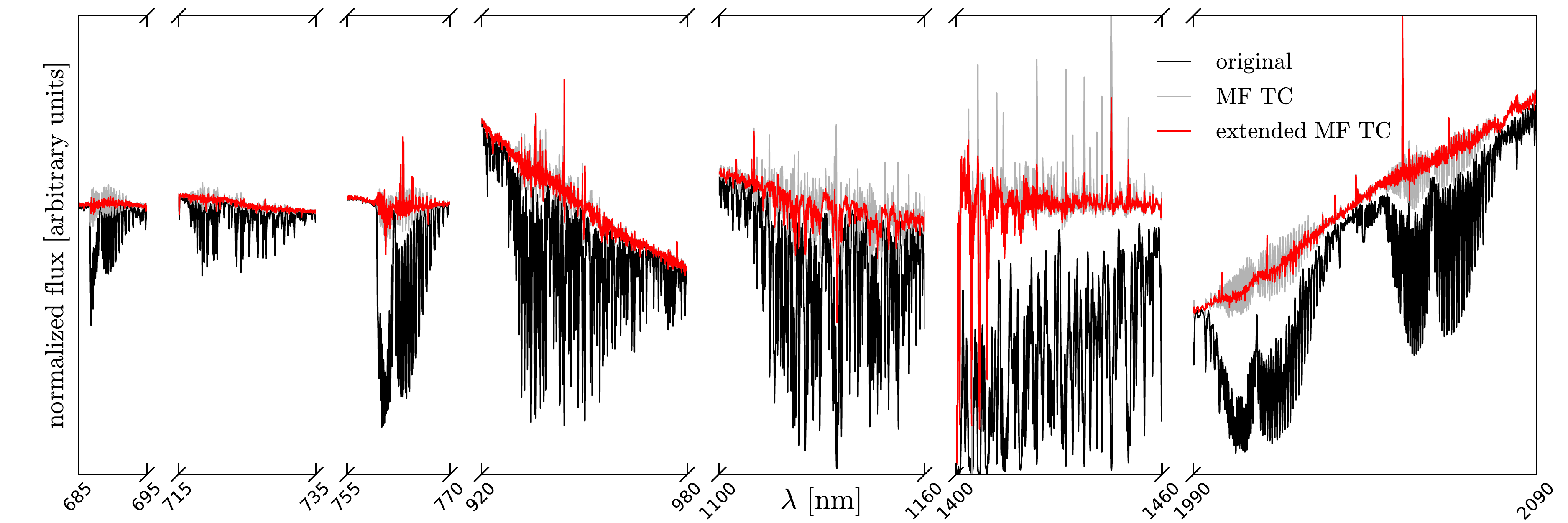}}
	{\includegraphics[width=0.9\textwidth]{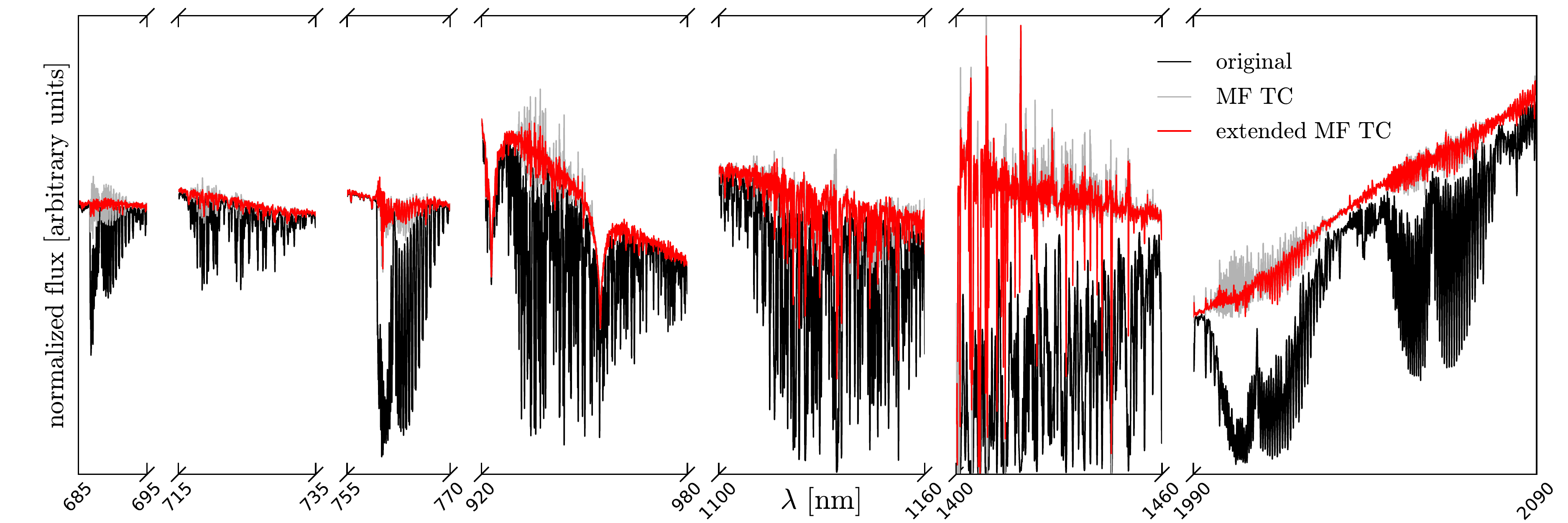}}
	\caption{Comparison between telluric correction with the classical \texttt{molecfit} approach (MF TC) and with the extended \texttt{molecfit} approach (extended MF TC). The upper star corresponds to a flux-standard star and the lower star to X0073 (HD 18769). The figure shows different telluric regions in the VIS arm and NIR arm. Within each of these regions, the corrected spectrum is normalized to unit mean. }
	\label{telluricPlot1}
\end{figure*}

We corrected the 1D extracted spectra for telluric absorption with \texttt{molecfit} \citep{Smette15,Kausch15}. \texttt{Molecfit} uses a radiative transfer code and a molecular line  database to calculate a synthetic spectrum of the Earth's atmosphere based on local weather conditions and standard atmospheric profiles. In addition, \texttt{molecfit} also models the instrumental line spread function. Once the characteristics of the instrument have been taken into account, the synthetic transmission spectrum can be used to correct an observed spectrum for telluric absorption.

\begin{table}[!h]
\small
	\centering
	\caption{Parameters of \texttt{molecfit} used to correct XSL spectra for telluric absorption. See \citet{Noll14} for an explanation of the meaning of the different parameters.}
	\label{tab:MFparameters}
	\begin{tabular}{ll} 
		\hline \hline
		Parameter & Value\\
		\hline
		 FTOL & 0.01 \\
		 XTOL & 0.01 \\
		 LIST\_MOLEC & O3 (UVB)\\
		 {} & H2O O2 O3 (VIS)\\
		 {} & H2O CO2 CO CH4 O2 (NIR)\\
		 FIT\_MOLEC & 0 (UVB)\\
		 {} & 1 0 0 (VIS)\\
		 {} & 1 0 0 0 0 (NIR)\\
		 RELCOL & 1.0 (UVB)\\
		 {} & 1.0 1.0 1.0 (VIS)\\
		 {} & 1.0 1.05 1.0 1.0 1.0 (NIR)\\		 
		 FIT\_CONT & 1 \\
		 CONT\_N & 3 \\
		 CONT\_CONST & Average of spectrum \\
		 FIT\_WLC & 1 \\
		 WLC\_N & 0 \\
		 WLC\_CONST & 0.0 \\
		 FIT\_RES\_BOX & 0 \\
		 KERNMODE & 0 \\
		 FIT\_RES\_GAUSS & 1 \\
		 RES\_GAUSS & 1.0 \\
		 VARKERN & 1 \\
		\hline
	\end{tabular}
	\normalsize
\end{table}

Table~\ref{tab:MFparameters} details the relevant parameters that we used for running \texttt{molecfit}. Most of these are taken from \citet{Kausch15}, who also applied \texttt{molecfit} to the processing of X-Shooter spectra. In the UVB arm we only corrected for absorption by $\mathrm{O}_3$, in the VIS arm we corrected for absorption by $\mathrm{H}_2\mathrm{O}$, $\mathrm{O}_2$ and $\mathrm{O}_3$, and in the NIR arm for $\mathrm{H}_2\mathrm{O}$, $\mathrm{C}\mathrm{O}_2$, $\mathrm{C}\mathrm{O}$, $\mathrm{C}\mathrm{H}_4$ and $\mathrm{O}_2$. Except for water vapor, the abundances of the different mole\-cular species are assumed to be constant and equal to the standard atmospheric values. The total amount of water vapor along the pointing direction of the observation is a free parameter of \texttt{molecfit} that can be fit together with the other free parameters. The wavelength regions used for the fitting procedure are specified in Table \ref{tab:WRegions}. For the NIR arm and the VIS arm, these regions are almost the same as those used in \citet{Kausch15}.

\begin{table}[!h]
	\centering
	\caption{Wavelength regions (air) used by \texttt{molecfit} for fitting atmospheric transmission spectrum. For each region, the most important molecule of the transmission spectrum is also given.}
	\label{tab:WRegions}
	\begin{tabular}{ccc} 
		\hline \hline
		Arm & Wavelength & Molecule\\
		{} & region [nm] \\
		\hline
		 UVB & 302-350 & O$_3$\\
		 \hline
		 VIS & 685-694 & O$_2$\\
		 VIS & 758-777 & O$_2$\\
		 VIS & 929-945 & H$_2$O\\
		 \hline
		 NIR & 1120-1130 & H$_2$O\\
		 NIR & 1470-1480 & H$_2$O\\
		 NIR & 1800-1810 & H$_2$O\\
		 NIR & 2060-2070 & CO$_2$\\
		 NIR & 2350-2360 & CH$_4$\\
		 \hline
	\end{tabular}
\end{table}

It is known that the wavelength calibration of X-Shooter is not perfect: \citet{Moehler15} reports shifts with an average peak-to-peak amplitude of 0.28 pixels, corresponding to 0.0042 nm and 0.0168 nm for our VIS and NIR data, respectively. 
This results in small offsets between the wavelength scale of the spectrum and the synthetic telluric transmission spectrum produced by \texttt{molecfit}. We alleviate this problem by applying \texttt{molecfit} in a slightly different way. First, we use the ``classical'' \texttt{molecfit} approach, as described in \citet{Smette15}, to derive the precipitable water vapor column (PWV). Then we cut the spectrum into different wavelength segments. The segments are chosen such that each one contains at least one telluric feature and, if possible, also a ``clean'' region which is (almost) free of telluric contamination. We fix the PWV parameter to the value found in the first iteration and apply \texttt{molecfit} to each of the wavelength segments. As a final step, the corrected wavelength segments are recombined.
Across regions where wavelength segments overlap, we use linearly progressive weights to average the two sets of data. 

\begin{figure}[!h]
 \centering
\includegraphics[width=.95\linewidth,trim=0 0 0 10cm,clip]{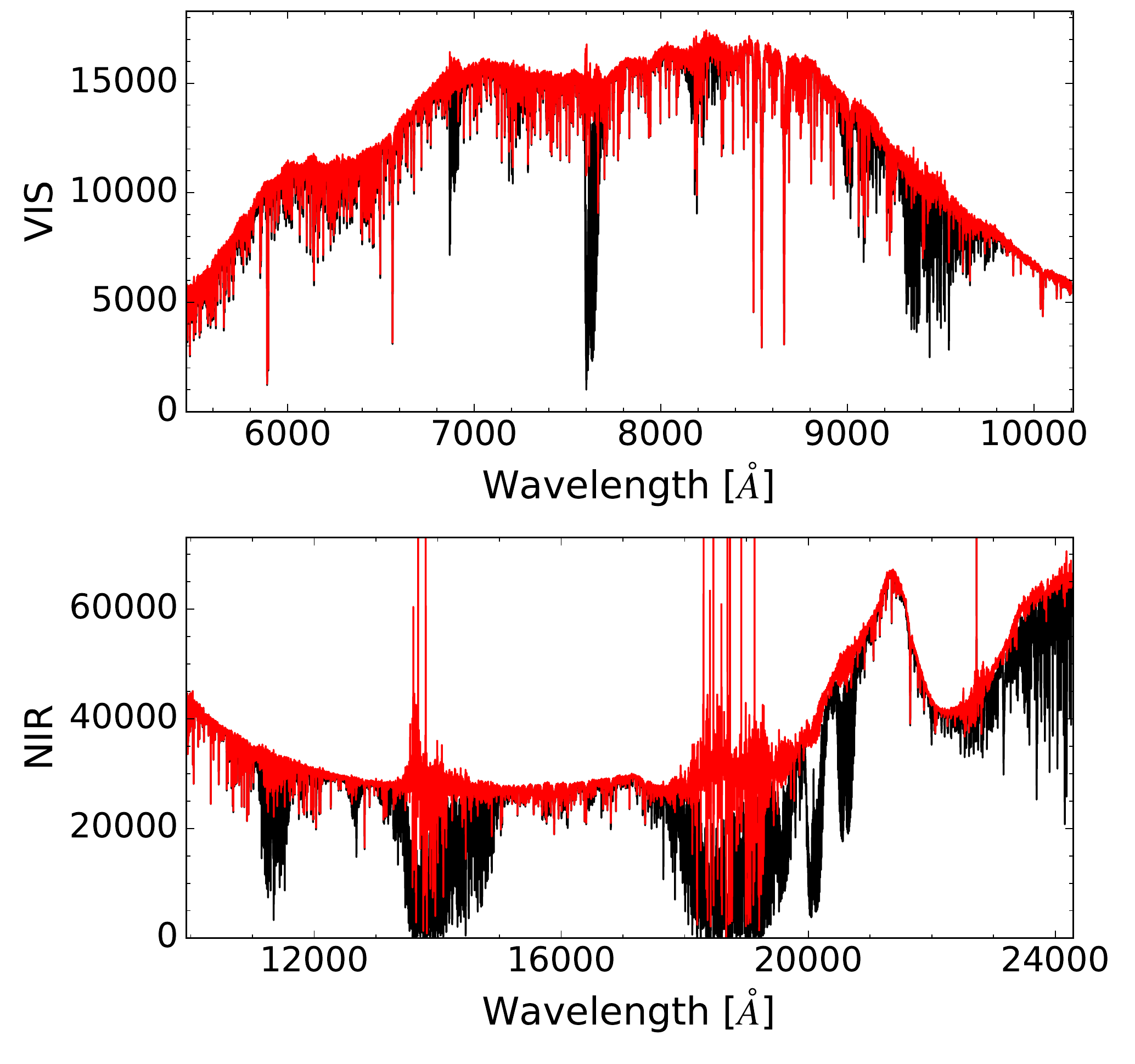}
  \caption{Before (black) and after (red) the telluric absorption correction process for X0365 (VIS and NIR arms). }
  \label{fig:tell_corr}
\end{figure}

Our ``extended \texttt{molecfit}'' approach, where the spectrum of a particular arm is divided into smaller wavelength segments, allows \texttt{molecfit} to find better local wavelength solutions and line spread functions. The final telluric-corrected spectra have a much smaller variance.
We use the extended \texttt{molecfit} approach to correct spectra in the VIS and NIR arms. The boun\-daries of the different wavelength segments are given in Table \ref{tab:WChunks}.
Figure~\ref{telluricPlot1} compares the two telluric correction approaches: the classical \texttt{molecfit} approach (in gray) and the extended \texttt{molecfit} approach (in red). 
Figure~\ref{fig:tell_corr} illustrates the results for the full VIS and NIR wavelength ranges.
For the UVB arm, we use the classical \texttt{molecfit} approach since for this arm the wavelength region that contains telluric lines is relatively small.

\begin{table}[!h]
	\centering
	\caption{Wavelength segments (air) used for the extended \texttt{molecfit} approach, as described in the text. The extended \texttt{molecfit} approach is only used for the VIS arm and the NIR arm.}
	\label{tab:WChunks}
	\begin{tabular}{cccc} 
		\hline \hline
		Arm & Wavelength & Arm & Wavelength \\
		{} & segment [nm] & {} & segment [nm]\\
		\hline
		 VIS & $\lambda_{\mathrm{start}}$-637 & NIR & $\lambda_{\mathrm{start}}$-1105\\
		 VIS & 633-712 & NIR & 1095-1185\\
		 VIS & 708-752 & NIR & 1175-1255\\
		 VIS & 748-782 & NIR & 1245-1305\\
		 VIS & 778-862 & NIR & 1295-1455\\
		 VIS & 858-926 & NIR & 1445-1605\\
		 VIS & 922-$\lambda_{\mathrm{end}}$ & NIR & 1595-1755\\
		 {} & {} & NIR & 1745-1880\\
		 {} & {} & NIR & 1870-1985\\
		 {} & {} & NIR & 1975-2040\\
		 {} & {} & NIR & 2030-2085\\
		 {} & {} & NIR & 2075-2285\\
		 {} & {} & NIR & 2275-2385\\
		 {} & {} & NIR & 2375-$\lambda_{\mathrm{end}}$\\
		 \hline
	\end{tabular}
\end{table}

\subsection{Atmospheric extinction correction}
\label{sec:atmos_scatter}

Before computing instrument response curves and applying them, the spectra of science and flux-standard stars were corrected for observation-specific effects. We corrected the frames for exposure time, gain and continuous atmospheric extinction as follows:
 \begin{equation}
F_{out} = \frac{F_{in}}{exptime \times gain}\, 
\times \,  10^{(0.4 \cdot airmass \cdot  extinct)}
\end{equation},
where $F_{in}$ is the flux of the observations as available after correction for telluric absorption bands. The airmass, exposure time (\textit{exptime}) and gain were taken from the observation header. 

The atmospheric extinction (\textit{extinct}, expressed in magnitudes per unit airmass) combines aerosol scattering (Mie diffusion, $k_{aero}$) and Rayleigh scattering ($k_{ray}$), following  \citet{Patat11}. Molecular absorption (including broad features that are sometimes dealt with like a contribution to the extinction continuum) were already accounted for by {\tt molecfit}.
We model the aerosol contribution as advocated by \citet{Moehler14}:
\begin{equation}
k_{aero} = 0.014 \times (\lambda[\mu m])^{-1.38}
\end{equation}.
The Rayleigh scattering mostly depends on the observation airmass, and only weakly on the atmospheric conditions. 
In the zenith direction, we model it as: 
\begin{equation}
k_{ray} = \frac{p}{1013.25} \times (0.00864+6.5 \times 10^{-6} H) \times \lambda^{-(3.916+0.074 \lambda+0.050/\lambda)} \end{equation},
with pressure $p$ $=$ 744 hPa, height $H$ $=$ 2.64 km and wavelength $\lambda$ in $\mu$m \citep[values from][]{Noll12}.

\begin{figure}[!h]
 \centering
\includegraphics[width=.95\linewidth,clip]{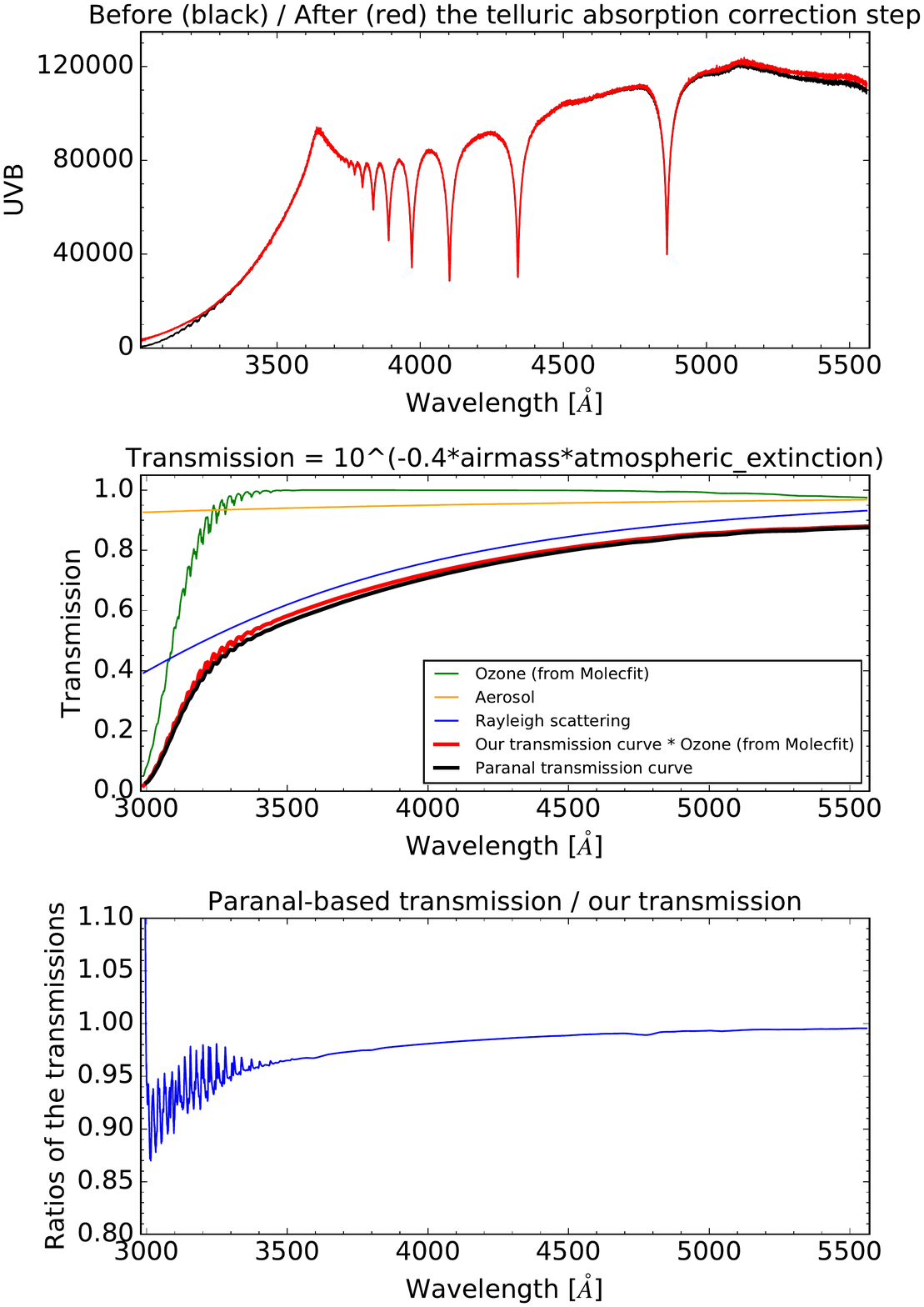}
  \caption{\textbf{Top:} UVB flux-standard spectrum (related to X0365) corrected for O$_3$ with Molecfit (red spectrum), compared to an uncorrected one (black spectrum). \textbf{Middle:} Our own transmission curve (red curve) -- with the details of its three components -- compared to the pipeline one (black curve). 
  \textbf{Bottom:} Ratios of the transmissions. }
  \label{fig:atmos_ext_curve}
\end{figure}

Figure~\ref{fig:atmos_ext_curve} compares our extinction curve to the XSH pipeline one. From the bottom panel of that figure, 
we can see that the main differences are in the ozone Huggins bands, which are not deep enough in the reference curve. 

\subsection{Response curve and first flux calibration}
\label{sec:response_curve}

In the XSH pipeline the 2D flat-field spectra are {\it not} norma\-lized across the surface of each grating order before they are divided into the program data. Hence the division corrects both the pixel-to-pixel variations of the sensitivity of the detector and major transmissions variations, such as those produced by the blaze functions of the gratings. This choice ensures that extracted single-order spectra within one arm are easily connectable. In exchange for this benefit, the arm-spectra (as illustrated in Figs. \ref{fig:order_by_order} and \ref{fig:tell_corr}) contain an imprint of the (inverted) spectra of the lamps used to illuminate the flat-field exposures. Unfortunately, the X-Shooter flat-field lamps are less stable than other elements of the acquisition chain, in particular in the NIR arm. Hence, when pairing observations of a science target and of a spectro-photometric standard star, it is recommended to use the same flat-field images for the two data sets\footnote{See \url{http://www.eso.org/observing/dfo/quality/XSHOOTER/qc/problems/problems_xshooter.html\#NIR_FF}; current recommendations extend to the VIS and UVB arms.}.

To determine a response curve from the 1D spectrum of a flux standard, 
we performed a $\chi^2$-minimization between that observation on one hand,
and the product of its theoretical spectrum  
and the unknown response curve on the other. The response curve is represented with a high-order spline polynomial.
The typical number of spline nodes is 35 for the UVB and NIR arms and 60 for the VIS.
These spline nodes are not all regu\-larly spaced; they are placed sparsely where spurious oscillations should be avoided (e.g., across the telluric water bands in the NIR arm) and more tightly where small-scale instrumental effects must be accounted for (e.g., features at 365 nm due to the combination of two different flat-fields in the UVB arm\footnote{\label{note_uvb_ff} \url{https://www.eso.org/observing/dfo/quality/XSHOOTER/pipeline/pipe_problems.html\#uvb_flux}}, the dichroic features at 560-620 nm\footnote{\label{note_dich}\url{https://www.eso.org/sci/facilities/paranal/instruments/xshooter/knownpb.html}}, a flat-field bump at 2100 nm; see \citealt{Moehler14} for more details).
Furthermore, we masked the Balmer lines in the UVB and VIS arms, the O$_2$ lines in the VIS and the deepest telluric regions in the NIR.



We used the response curves for a first flux-calibration of the science spectra both for the narrow and wide slit observations.
The resulting intensities are physical flux densities, in erg s$^{-1}$ cm$^{-2}$ \AA$^{-1}$. These spectra are cleaned from the atmospheric and instrumental effects, {\it except} for the slit losses. The narrow-slit spectra, in particular, still do not reflect the absolute distribution of the stellar flux.


\subsection{Flux-loss correction and final energy distributions}
\label{sec:flux_loss}

For 85\% of the spectra, wide-slit observations of sufficient qua\-lity are available, and thus the narrow-slit spectra are corrected for wavelength-dependent flux-losses.
To this end, we computed a smoothed version of the ratio of the wide-slit over the narrow-slit spectrum, and then fit a low-order polynomial (typi\-cally of order 1 for UVB/VIS frames, and 3 for NIR frames).
Finally, we multiplied our narrow-slit spectra  by this polynomial. 
In a few cases (3 observations of 2 a priori non-variable stars, namely X0194 [HD~193281] and X0306/X0311 [HD~29391]), when only one of several observations of the same star had an associated wide-slit spectrum or an otherwise more reliable continuum, that observation was used as a reference for the other observation(s).



\medskip

\begin{figure*}[ht!]
 \centering
 \includegraphics[width=.8\textwidth,clip]{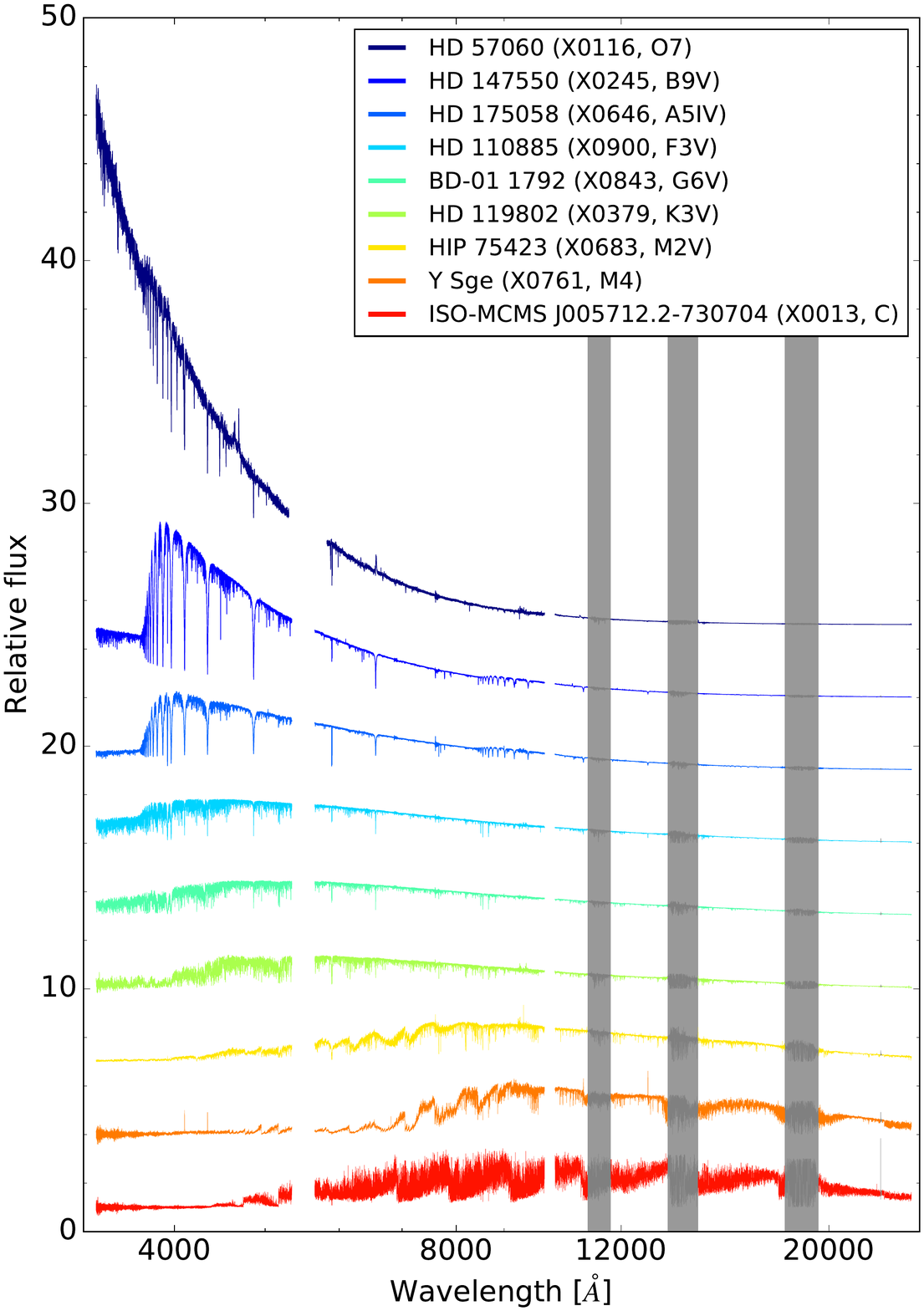}
  \caption{Typical XSL spectra in the OBAFGKMC sequence in log-scale of wavelength. For each star, the corresponding XSL observation number and spectral type are indicated in brackets. The gaps indicate the overlap regions between consecutive arms. Gray bands mask the deepest telluric regions in the near-infrared arm, as well as a few other local artifacts.}
  \label{fig:xsl_sequence}
\end{figure*}

In summary, our data reduction procedure uses the ESO pipeline until the 2D spectra are rectified and wavelength-calibrated. Subsequent steps were adjusted, mainly to improve the usage of pixel-flags, the sky subtraction, the telluric correction and the combined usage of narrow-slit and broad-slit observation for flux calibration. 85\,\% of the DR2 spectra are flux calibrated in the sense that they are corrected for all stable transmission factors of the acquisition chain, and for wavelength-dependent losses due to the narrow widths of the slits that provide the desired spectral resolution. Residual flux calibration errors may come from changes in telescope+instrument+sky transmission-curves between the observations of the science targets and of the spectrophotometric standard star, and from resi\-dual slit losses in the broad-slit observations. They are assessed via compa\-risons with external data sets in Section \ref{sec:phot}. A few examples of final DR2 spectra can be found in Fig.\,\ref{fig:xsl_sequence}.

\section{Properties of the DR2 spectra }
\label{sec:qa}

In this section we present an assessment of the overall properties of the spectra, including the quality of the wavelength and flux calibration, and properties of the line-spread function. We  
describe the quality flags associated with each published spectrum. 


\subsection{Typical signal-to-noise ratios}

\begin{figure}[!h]
 \centering
  \includegraphics[width=.85\linewidth,clip,page=1]{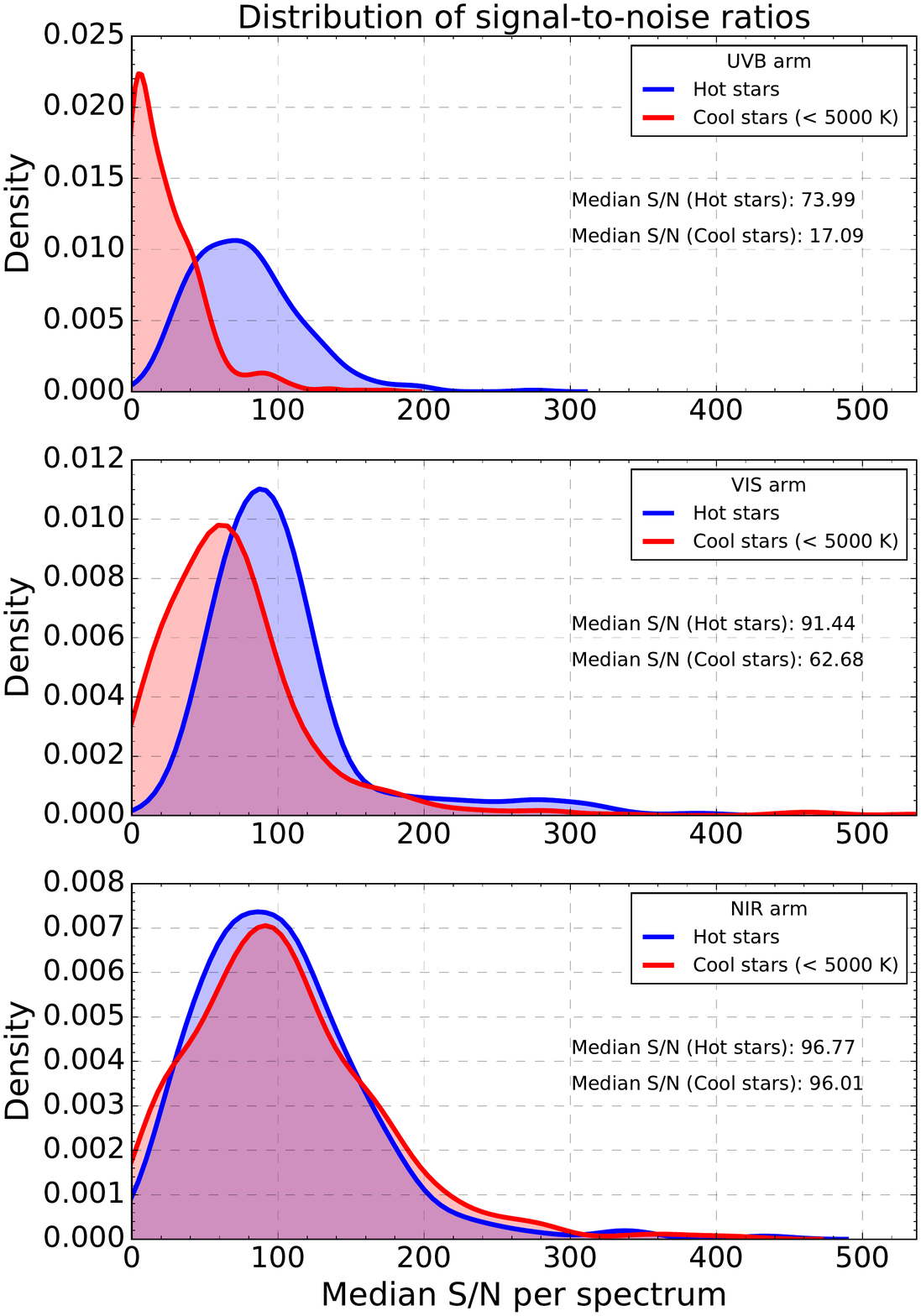}
  \caption{Distribution of the signal-to-noise ratio per arm.}
  \label{fig:histo_snr}
\end{figure}

Figure~\ref{fig:histo_snr} shows the distribution of the median signal-to-noise ratios (S/Ns) per pixel for our observations. 
Except for the coolest stars, typical S/N values are about 70, 90 and 96 for the UVB, VIS and NIR arms respectively. In fact, the S/N for cool stars (\teff $<$ 5000\,K) varies drastically with wavelength across the UVB arm. S/N similar to those of warmer stars are found in a temperature-dependent range of wavelength toward the red end of that arm.

Correlations between the errors in neighboring pixels are expected as a result of the redistribution of pixel fluxes when the original 2D-images are transformed into rectified order-images. They depend on the observation mode, on the seeing and on wavelength in a complex fashion. We have not attempted to cha\-racterize them in detail. 
In general however, they would extend over a few pixels, on a wavelength range similar to the (arm-dependent) resolution element of the spectra.

\subsection{Spectral resolution and line spread function}

We have determined the spectral resolution and the line spread function for each arm with the Penalized PiXel-Fitting method ({\tt pPXF} , \citealt{Cappellari2004, Cappellari2017}), using a subset of the PHOENIX theore\-tical spectral library \citep{Husser2013} as templates. The available PHOENIX library samples effective temperatures with a step of 200\,K (or of 500\,K above 12000\,K), surface gravities with a step of 0.5\,dex, and \feh with a step of 0.5\,dex (or of 1\,dex where \feh$<-1$).
The subset retained contains the 388 theoretical stars that are closest to the XSL library stars in the stellar parameter space, based on the stellar parameters derived from XSL observations by \citet{Arentsen19}. 

All XSL UVB, VIS and NIR spectra were fit one 200 \AA{}-wide wavelength region at a time. In the NIR, the telluric-contaminated regions (13\,500--14\,200 \AA{} and 17\,700--21\,000 \AA{}) were not analyzed.


\subsubsection{The spectral resolution}

The spectral resolution is described by the velocity dispersion measurements of the {\tt pPXF}  fit, as the template spectrum has a much higher resolution \citep[$R\sim 500\,000$,][]{Husser2013} than the X-Shooter spectra ($R \sim 10\,000$). As the observations span multiple observing periods, the measurements for stars from different observing periods were analyzed separately and then combined. 
In each observing period, the velocity dispersion results from the {\tt pPXF}  fitting were sigma-clipped with a 5-$\sigma$ threshold.
Sigma-clipping was itera\-ted until convergence was achieved.

\begin{figure*}[!ht]
 \centering
\includegraphics[width=.96\linewidth,clip]{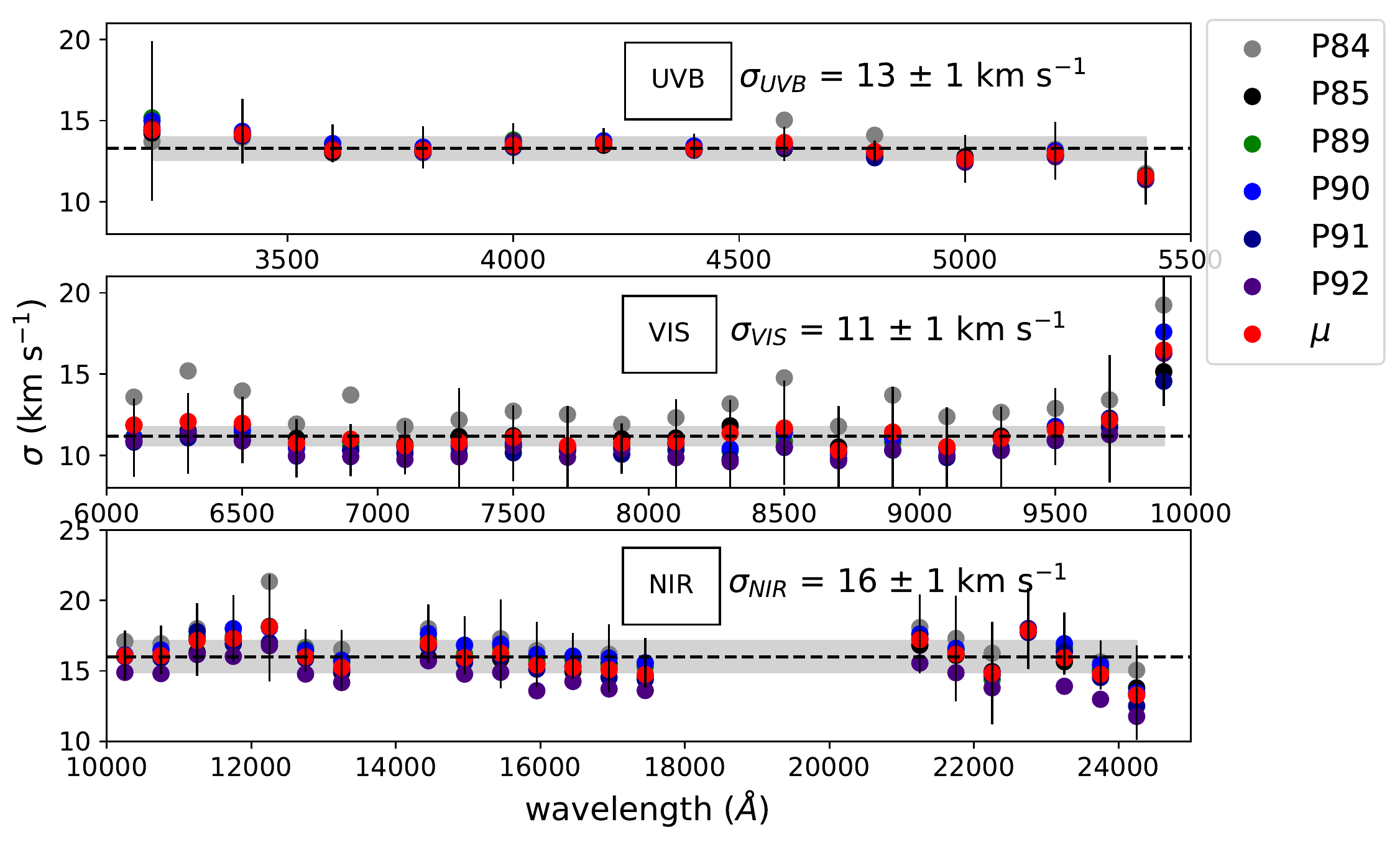}
  \caption{Spectral resolution wavelength dependence for each observing periods. Gray, black, green, blue, dark blue and indigo markers show the medians of measurements of data from observing periods ``P84'', ``P85'', ``P89'', ``P90'', ``P91'' and ``P92'' respectively. The red point shows the weighted average of all observing periods.  The error bars belong to period P90 measurements and are shown for illustrative purposes.}
  \label{fig:resolution_obsperiods}
\end{figure*}

Figure~\ref{fig:resolution_obsperiods} shows the wavelength dependence of the resolution for each observing period and X-Shooter arm. The $1~\sigma$ error bars belong to period P90 measurements, but are similar for all observing periods. The scatter around the median value comes from stars with lower S/N or a low number of spectral lines in each wavelength region. 
The red points show the weighted averages of the medians of all observing periods. The resolution is stable throughout the obser\-ving periods, with only P84 observations showing slightly lower resolution. The resolution $R=\lambda/\Delta \lambda$ is also relatively constant with wavelength, although \citet{Chen14} initially showed otherwise for the UVB arm. However, they only used the F, G, K stars of observing periods P84 and P85. The adopted spectral resolution for each arm is the average value over all wavelengths: $13 \pm 1$ \kms for UVB, $11 \pm 1$ \kms for VIS, and $16 \pm 1$ \kms for NIR. These values are shown with the black dashed lines and the $\pm1~\sigma$ gray shaded areas in Figure~\ref{fig:resolution_obsperiods}.

The spectral resolution measures are converted from $\sigma$ to FWHM assuming Gaussian line profiles in velocity space (FWHM\,$=2.35\sigma$) and used to re-evaluate the resolving power $R = c/FWHM$ ($c$ being the speed of light).
Our adopted values are close to the nominal ones\footnote{\url{https://www.eso.org/sci/facilities/paranal/instruments/xshooter/inst.html}}, as can be seen in Table~\ref{tab:resolution}. 

\begin{table}[!h]
\caption{Adopted spectral resolution values per arm}
\centering
\resizebox{\linewidth}{!}{%
\begin{tabular}{l c c r r}
\hline\hline
Arm & $\sigma$ & FWHM  & R  & R \\
 & $[\kms]$ & $[\kms]$ &  (our values) & (nominal values)\\
\hline
UVB & 13 & 30.61    & 9\,793    & 9\,700 \\
VIS & 11 & 25.90    & 11\,573   & 11\,400\\
NIR & 16 & 37.68    &  7\,956   & 8\,100\\
\hline 
\end{tabular}
}
\label{tab:resolution}
\end{table}

\subsubsection{The shape of the line spread function}

The deviations of the line spread function (LSF) from a Gaussian are described with Gauss-Hermite velocity moments $h_3$ and $h_4$, which are related to skewness and kurtosis respectively. These moments can be obtained with {\tt pPXF}  only when the signal-to-noise ratio is excellent. Therefore, $h_3$ and $h_4$ measurements from all observing periods were analyzed together. 
We computed the S/N in each wavelength bin 
(window of 200$~\AA$ each) and rejected measurements from spectra with S/N $<$ 200. Additionally, measurements from the very blue wavelengths, $\lambda<3600$~\AA{}, were not included due to low signal. The Gauss-Hermite moments were finally computed from 6510 highest-quality {\tt pPXF}  measurements. Table \ref{tab:lsf} shows the median and the standard deviation of the results per arm. To within one standard deviation, the line spread function in all arms is a Gaussian. 

\begin{table}[!ht]
\caption{Line-spread function parameters $h_3$ and $h_4$}
\centering
\resizebox{0.75\linewidth}{!}{%
\begin{tabular}{l r r r}
\hline\hline
 & 	UVB & VIS & NIR \\
\hline
$h_3$ & $0.00~(0.06)$ & $-0.03~(0.06)$ & $0.03~(0.05)$ \\
$h_4$ & $-0.03~(0.09)$ & $-0.01~(0.07)$ & $-0.01~(0.06)$\\
\hline 
\end{tabular}
}
\label{tab:lsf}
\end{table}


\subsection{Rest-frame velocity correction}
\label{sec:arm-velo}

The DR2 data are corrected for radial velocity and provided in the rest-frame (with wavelengths in air). The barycentric radial velocities, $cz$, were estimated separately in each of the three arms, because differences between arms are common and often in excess of 10\,\kms \citep[see][for a discussion]{Moehler15}. The $cz$ values are available in the header of the final FITS files, 
together with the barycentric corrections we applied\footnote{The barycentric radial velocity is the sum of the topocentric velocity and the barycentric correction.}.  

In the UVB and VIS arms, the final estimates were obtained with the ULySS package \citep{Koleva2009} and the ELODIE 3.2 spectral interpolator \citep{Wu2011}, as also used in the study of stellar parameters of \cite{Arentsen19}. 
This full-spectrum fitting approach requires an approximate a priori knowledge of the radial velocity for the method to converge, because it performs a local minimization using a Levenberg-Marquardt algorithm. The accuracy of the $cz$ guess should be better than a few times the FWHM of the spectral lines. 
At the resolution of XSL, the FWHM of the lines is on the order of 30 \kms, and since the barycentric velocities of the Galactic and LMC/SMC XSL stars are expected to span a range of up to about 400 \kms, we repeated the fit starting with a series of $cz$ guess in the interval from $-300$ to $+300$ \kms with a step of 100 \kms.
 For each value of the guess, the atmospheric parameters \teff, \logg and \feh as well as $cz$ and the broadening were re-adjusted. The solution with the lowest $\chi^2$ was adopted.

In the NIR, the adopted values of $cz$ were estimated by cross-correlation with the nearest model in the library of synthetic spectra of \citet{Husser2013}. Starting from the estimated parameters of \citet{Arentsen19}, we used $\chi^2$-minimization to identify an adequate near-IR representation of each near-infrared spectrum. The spectral ranges used for the cross-correlation were adjusted based on the effective temperature and spectral type, in order to include a sufficient number of spectral features. In the coolest stars, we avoided regions dominated by a single steep molecular bandhead, as the resulting $cz$ then depend on the way the spectra are renormalized to remove the dominant slope, before the cross-correlation.

For some spectra of cool stars, it was not possible to determine $cz$ in the UVB, or even in the VIS, because of an insufficient flux. When a measurement in the NIR was available, we then simply assigned the latter to the UVB and VIS.

During the preparation of DR2, several methods other than those finally adopted were employed to re-estimate the radial velocities of various data subsets. The distributions of velocity differences between methods typically display an approximately Gaussian peak, with a standard deviation of 2 to 3 \kms; non-Gaussian tails of the distributions typically contain $\sim$13\%, $\sim$18\% and $\sim$10\% of the stars in the UVB, the VIS and the NIR arms. Hence the value of 2 to 3 \kms is representative of the uncertainties on the rest-frame wavelengths of the DR2 data.


\subsection{Quality assurance}

After the data reduction, we assigned a number of quality flags to our data based on visual inspection and on experience gained by comparing the observations with synthetic or other empirical spectra
\citep[e.g.,][]{Gonneau17, LanconIAU18, Arentsen19}. Table~\ref{table:key_dico} (item ``Quality flags (per arm)'') describes our flags. 
The flags were added to the headers of the final FITS files, and peculiarities specific to individual spectra are listed in Table\,\ref{tab:peculiar}.
We note that this list is not exhaustive, as some artifacts might have been overlooked.

More generic caveats concerning the data are the following.
In the UVB, the spectra tend to be wavy near 0.35$\mu$m (the combination of two different flat-fields D2 and QTH in this range leads to errors, see footnote \ref{note_uvb_ff}) and between 0.545 -- 0.55$\mu$m (known instabilities of the transmission of the dichroic plate that feeds the UVB and VIS arms of X-Shooter, see footnote \ref{note_dich}). 

For the VIS arm, frequent artifacts are seen in the spectral shape between 0.54 -- 0.59$\mu$m (dichroic issue, see footnote \ref{note_dich}). 
The continuum at the red end, beyond 0.97$\mu$m, can also have some artifacts (second dichroic region; effect $>5\,\%$). Molecfit corrects only for molecular absorption lines, hence telluric absorption by me\-tals is not removed (an example is the frequent contamination of the stellar NaI doublet by telluric Na, between 588 and 592 nm).

In the NIR, a step sometimes arises between the two reddest orders (near 2.25$\mu$m). The noise spectrum does not capture the 1.1$\mu$m telluric band sufficiently strongly. The slope of the spectrum is sometimes poorly determined below 1.05$\mu$m (dichroic region; effect $>5\,\%$). The $K-$band is usually of poor quality for hot stars (K-band contiuum wavy, lower S/N). Some of these stars are flagged with HAIR\_NIR (meaning that there are some narrow spikes in the NIR spectrum, cf Table~\ref{table:key_dico}).


\subsection{Photometric comparison}
\label{sec:phot}

\citet{Chen14} compared the DR1 (UVB and VIS) spectra with literature spectra taken from NGSL \citep{Gregg06}, the calcium-triplet library CaT \citep{Cenarro01}, and also with higher-resolution spectral libraries such as UVES POP \citep{Bagnulo03} and ELODIE \citep{Prugniel01}. They found a good agreement in the line shapes and depths between XSL and these libraries.

Here we focus on the broad-band spectral energy distribution across
the single-arm spectra of DR2. Based on synthetic photometry, we compare colors within the UVB and VIS arms to those of the MILES library \citep{Sanchez06}; in the near-infrared, we confront the XSL colors with those of the 2MASS survey \citep{Cohen03, 2mass06}, and with synthetic colors of the IRTF/E-IRTF spectra \citep{Rayner09, Villaume17}.


\subsubsection{Comparison with MILES}
\label{sec:miles}

\begin{figure}[!h]
    \centering
\includegraphics[height=.78\linewidth,angle=270,trim=4cm 4cm 3cm 5.5cm,clip]{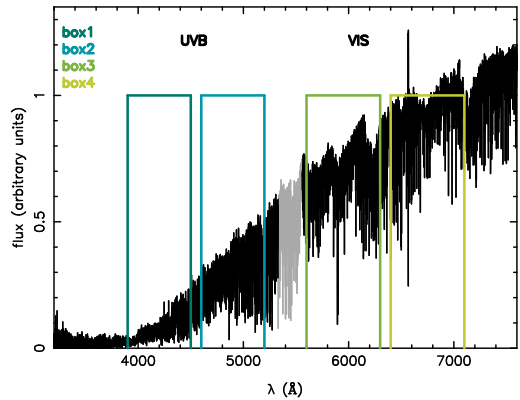}
    \caption{Filters designed for the comparison with MILES spectra, overlaid on the XSL spectrum  of a cool star.
    The filters avoid the spectral regions most strongly affected by the unstable dichroic between the UVB and VIS arms (plotted in gray).  }
    \label{fig:filters_miles}
\end{figure}

Many of the XSL stars were initially selected from the MILES collection \citep{Sanchez06}. Forty MILES stars were present in XSL DR1, and \citet{Chen14} found a good agreement between these data in the UVB and VIS arms. In the present release, 173 spectra of 142 stars have a MILES counterpart, of which 164 are corrected for slit-losses. To avoid complications due to the edges of the X-Shooter arms, we define artificial broad-band filters to compare the synthetic photometry of the two data sets (Table~\ref{tab:filters_waves} and  Figure~\ref{fig:filters_miles}). 

\begin{table}[!h]
\caption{Artificial filters used in Section \ref{sec:miles}.}
\centering
\begin{tabular}{l l r r}
\hline\hline
Reference set & Filter & $\lambda_{min} [\AA]$ & $\lambda_{max} [\AA]$ \\
\hline
MILES & box1 &  3\,900 & 4\,500 \\
& box2 & 4\,600 & 5\,200 \\
& box3 &  5\,400 &  6\,200 \\
& box4 & 6\,300  &  7\,100\\
\hline
\hline 
\end{tabular}
\label{tab:filters_waves}
\end{table}

\begin{figure}[!h]
    \centering
  \includegraphics[height=.82\linewidth,angle=270, trim=1cm 4cm 0.5cm 5cm, clip]{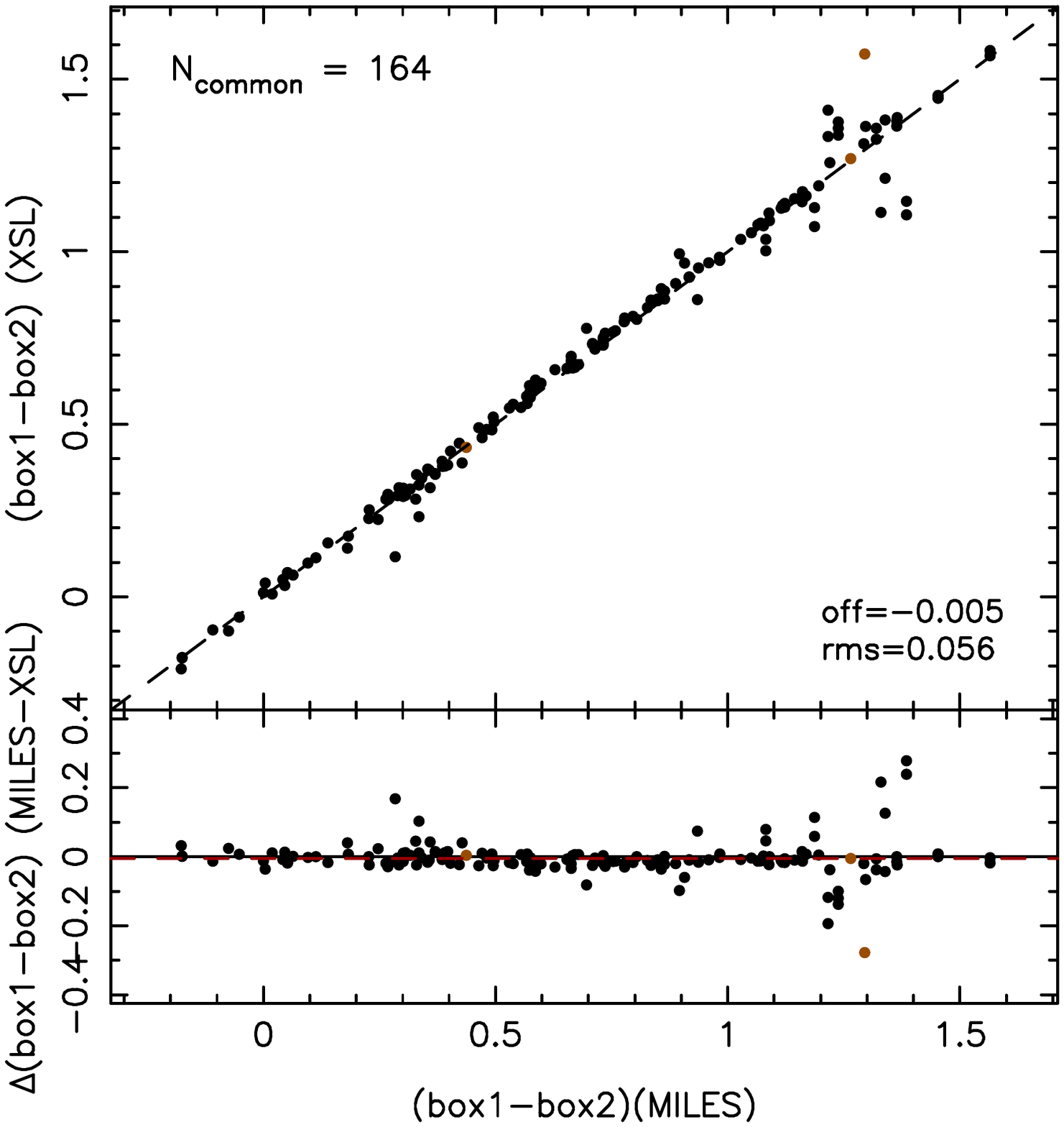}\\
  \includegraphics[height=.82\linewidth,angle=270, trim=1cm 4cm 0.5cm 5cm, clip]{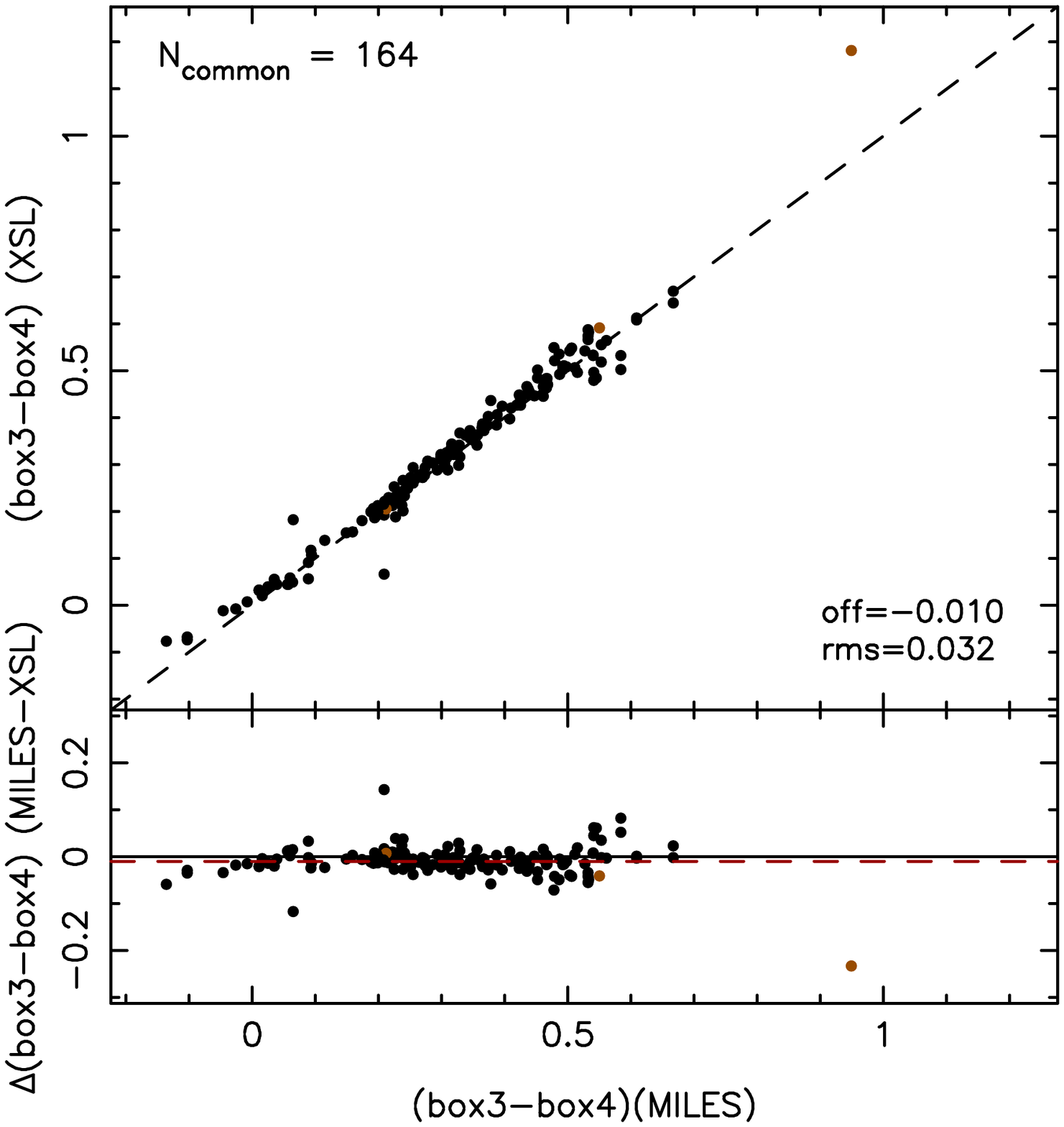}
    \caption{Comparison of synthetic colors between XSL and MILES. Only XSL-spectra with slit-loss corrections are used.}
    \label{fig:color_miles}
\end{figure}

Figure~\ref{fig:color_miles} shows how the MILES and XSL synthetic colors compare. We find excellent agreement in general, with a mean offset of $-0.005/-0.01$ mag and an rms scatter of 0.056/0.032 mag, respectively for colors $(box1-box2)$ and $(box3-box4)$.
We have examined outliers with color differences larger than 0.1\,mag individually, using plots of the spectra and the extensive comparison of XSL spectra with the theoretical spectra of \citet{Husser2013} that is the subject of another article (in prep.). The clear outlier at the red extreme in the bottom panel is BD+19 5116B (XSL observation X0440, spectral type M4.0Ve), a flare star. We observed this star at a different phase than MILES. 
The star is also an outlier in ($box1-box2$). Six other outliers at the red end of ($box1-box2$) are cluster stars. Five of these are in NGC\,6838. We trust their XSL energy distributions to better than 10\,\% because the comparison of these spectra with reddened theoretical models raises no alarms and leads to an extinction estimate that agrees with the values listed in \citet{DiCecco15} for the cluster (article in preparation). One object is in NGC\,1904, and again best-fit models are not alarming and agree with the very small reddening toward that cluster found in the literature. Still in ($box1-box2$), the bluest outlier is a hot star, which is also represented very well with models. Remo\-ving those outliers, the rms scatter in ($box1-box2$) drops from 0.056 to 0.030\,mag. The two outliers in ($box3-box4$) not yet discussed are the hot star just mentioned, and an F-type star for which the fit with model spectra excludes large error on the slope in the VIS arm. Removing these, the rms scatter for ($box3-box4$) drops from 0.032 to 0.022\,mag. Table~\ref{tab:DR2photom} summarizes these dispersion measurements.

\begin{table}[!h]
\caption{Standard deviations of the color-differences between XSL and external datasets}
\centering
\begin{tabular}{lccl}
\hline \hline
External & Color  &  Std. dev. & Comments \\
reference &  &  $\sigma$ [mag]   & \\
\hline
MILES & ($box1-box2$) & 0.030 & 0.056 incl. 7 outliers \\
MILES & ($box3-box4$) & 0.022 & 0.032 incl. 3 outliers \\
2MASS \rule[0pt]{0pt}{12pt} & $(J\,-\,H)$ & 0.032 & 0.050 incl. 2MASS errors \\
2MASS & $(H\,-\,K_s)$ & 0.038 & 0.056 incl. 2MASS errors \\
IRTF \rule[0pt]{0pt}{12pt} &  $(J\,-\,H)$ & 0.023 & 0.033 incl. 6 outliers \\
IRTF & $(H\,-\,K_s)$ & 0.019 & 0.029 incl. 4 outliers \\
\hline \hline
\end{tabular}
\label{tab:DR2photom} \\
Note: The outliers are not the same for all colors. The 2MASS standard deviations listed each include six 4-$\sigma$ outliers. See text for details. 
\end{table}


\subsubsection{Comparison with 2MASS}
\label{subsec:SDSS2MASS}

\begin{figure}[!h]
    \centering
  \includegraphics[height=.78\linewidth,angle=270,trim=4cm 4cm 2.5cm 5.5cm,clip]{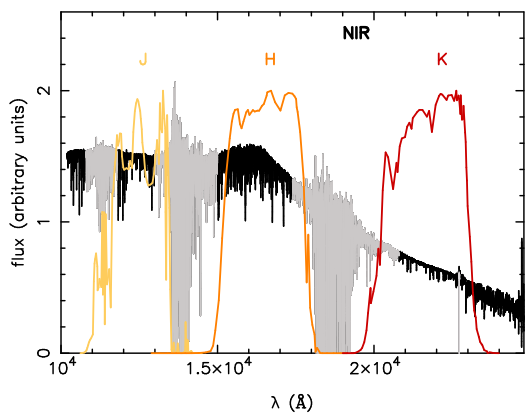}
    \caption{Response curves of the 2MASS filters, 
    overlaid on an XSL spectrum. 
    The transmission curves are taken from     \citet{Cohen03}. } 
    \label{fig:filters}
\end{figure}

The 2MASS filters $J$, $H$ and $K_s$ 
fit entirely into the wavelength range of the 
NIR arm of X-Shooter, as seen in Figure \ref{fig:filters}.
We compute synthetic photometry compatible with 2MASS using the relative spectral response curves and reference fluxes of \citet{Cohen03}.
The 2MASS filters $J$, $H$ and $K_s$ 
fit entirely into the wavelength range of the 
NIR arm of X-Shooter, as seen in Figure \ref{fig:filters}).

\begin{figure}[!h]
    \centering
    \includegraphics[height=.82\linewidth,angle=0,
     clip,trim=5cm 0.5cm 6cm 1cm]{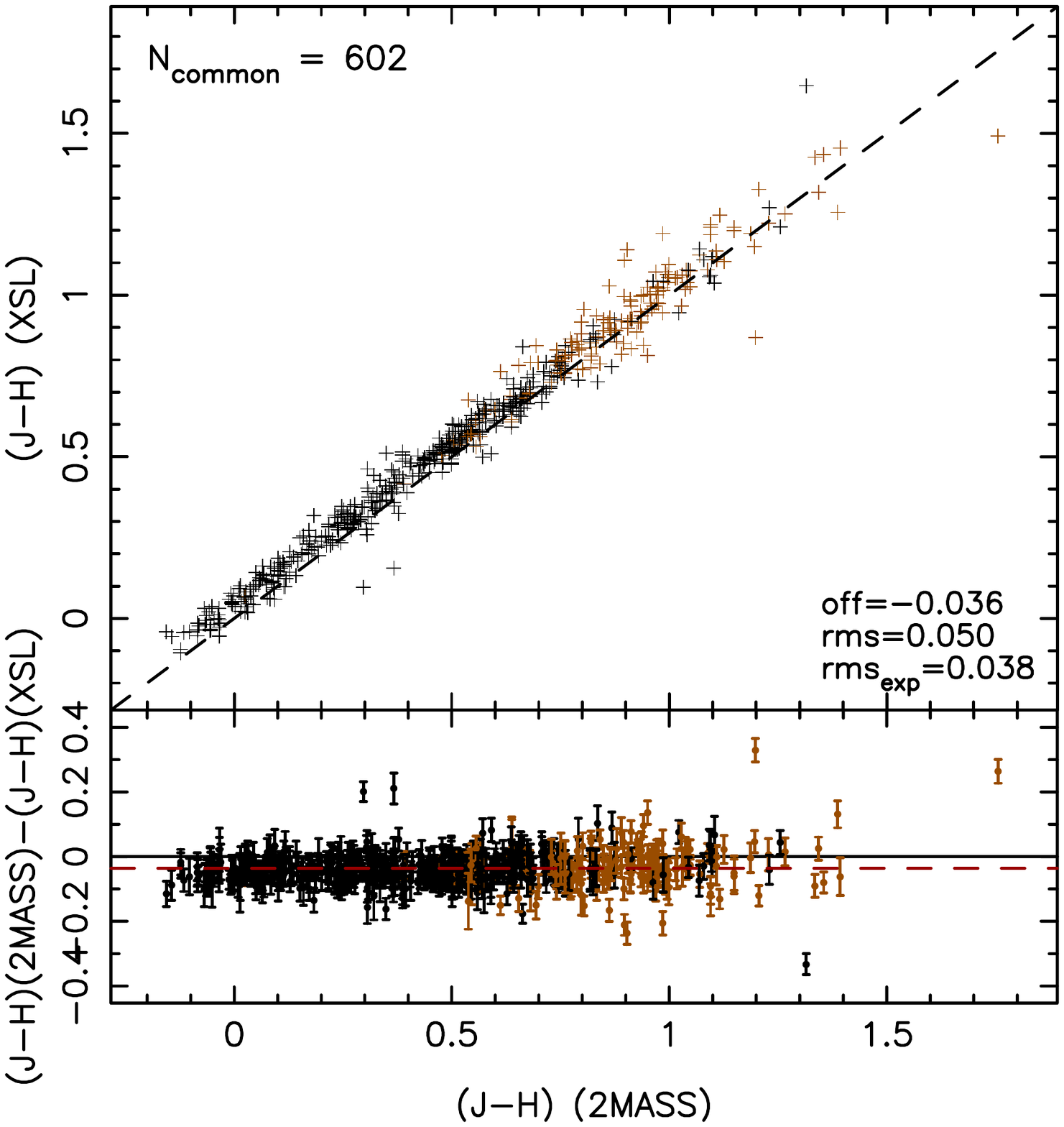}\\
    \includegraphics[height=.82\linewidth,angle=0,
    clip,trim=5cm 0.5cm 6cm 1cm]{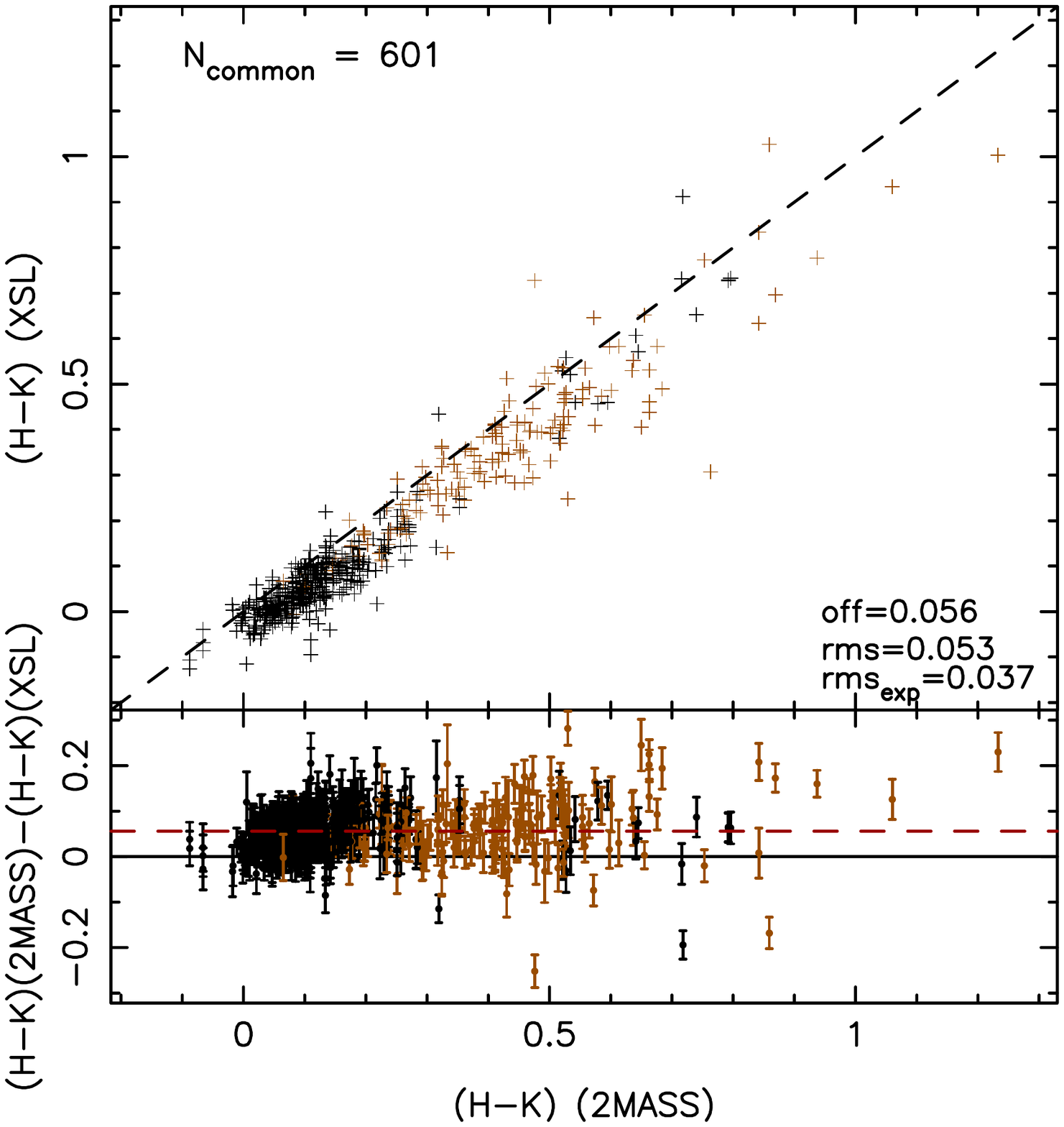}
    \caption{Comparison of synthetic colors between XSL and 2MASS. Only XSL-spectra with slit-loss corrections are used. The brown points correspond to cool stars ($T_{\mathrm{eff}}<3500\,$K or spectral type M and C. The legends provide the average offset between the colors of the two data sets, the rms scatter around the mean, as well as the part of the rms scatter that is explained by the 2MASS error bars shown in the bottom panels.).}
    \label{fig:color_2mass}
\end{figure}

Figure~\ref{fig:color_2mass} shows the results of the comparison.
Overall there is good agreement. Small systematic
offsets indicate small zero point errors in the exercise; they are in opposite directions for $(J-H)$ and $(H-K_s)$,
suggesting a normalization difference in the $H$ band \citep[as also mentioned in][]{Villaume17}. 
We find an rms scatter around the mean of 0.050\,mag for $(J-H)$ and 0.053\,mag for $(H-K_s)$.
After removing the contribution explained by 2MASS errors in quadrature, the residual scatter is
of 0.032\,mag in $(J-H)$ and 0.038\,mag in $(H-K_s)$. These are still conservative estimates of the errors in the near-infrared XSL colors, which are clarified in Section~\ref{sec:IRTF}.
In both panels, the larger dispersion toward redder objects is due to cool stars (M and C stars, and stars with $\teff < 3500 $K), of which many are known or suspected variables.


\subsubsection{Comparison with IRTF}
\label{sec:IRTF}

\begin{figure}[!h]
    \centering
\includegraphics[height=.82\linewidth,angle=270,trim=1cm 4cm 0.5cm 5cm,clip]{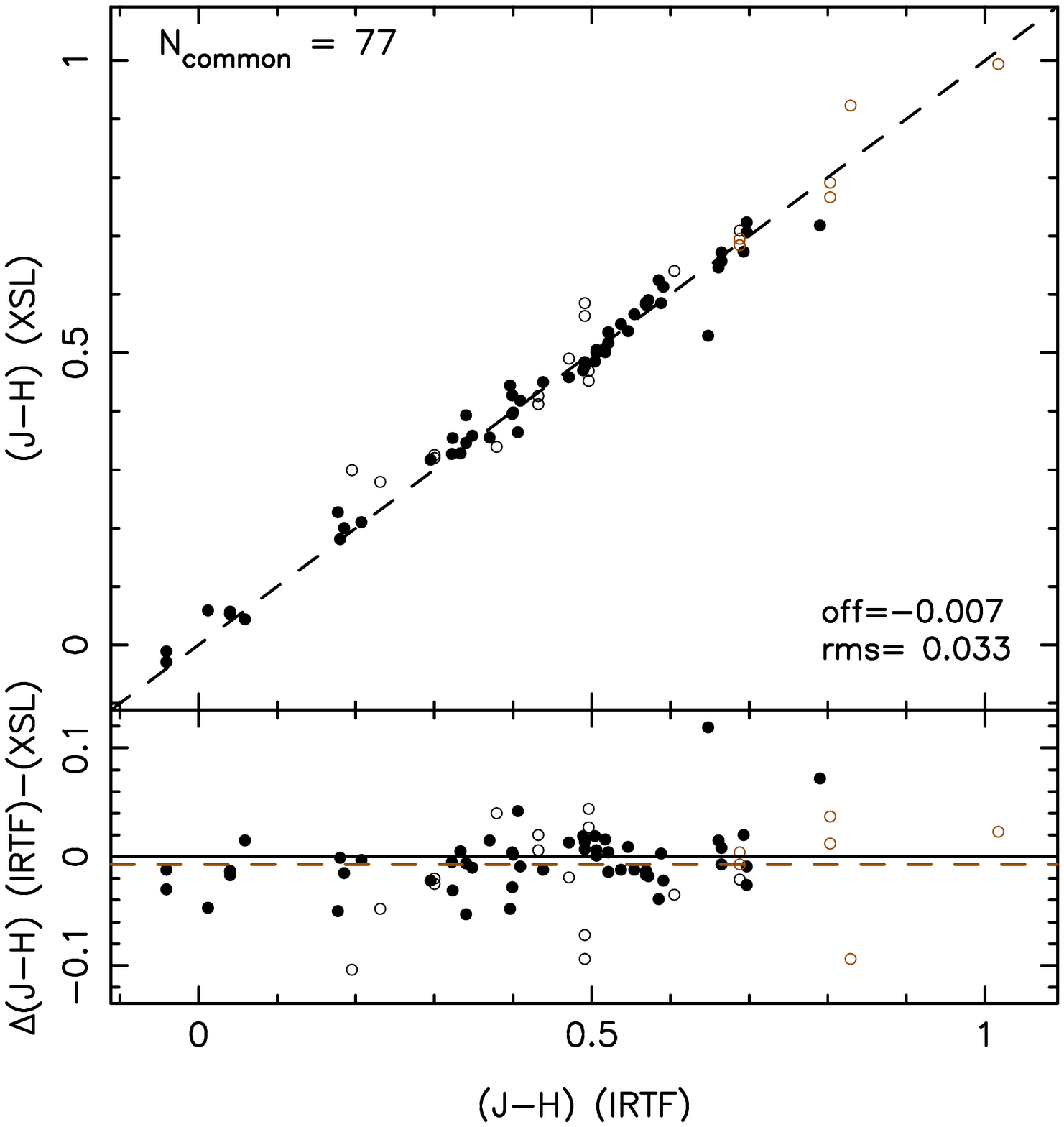} \\
\includegraphics[height=.82\linewidth,angle=270,trim=1cm 4cm 0.5cm 5cm,clip]{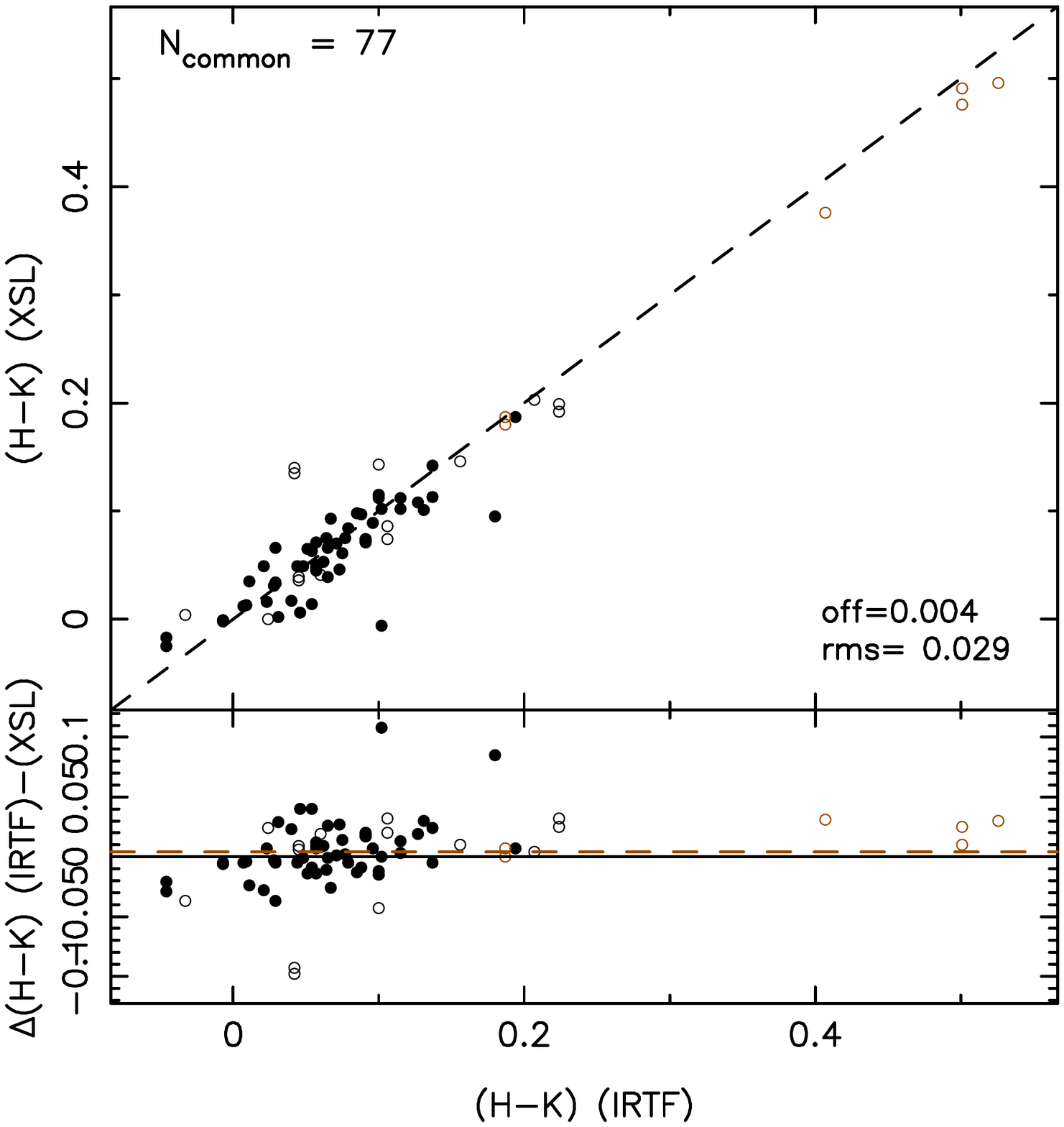}
    \caption{Comparison of synthetic colors between XSL and IRTF (open circles for IRTF and filled circles for E-IRTF). The brown symbols correspond to cool stars (M/C stars, and stars with no spectral type and $\teff < 3500 $ K). 
    }\label{fig:color_irtf}
\end{figure}

Empirical near-infrared spectral libraries are still rare in the lite\-rature.
We choose to compare XSL with the IRTF spectral library \citep{Rayner09} 
and its later extension \citep[hereafter E-IRTF,][]{Villaume17} as their wavelength range overlaps with ours (0.7 -- 2.5 $\mu$m) and the spectra are flux-calibrated. Their resolution is $R \sim$\,2000, and their signal-to-noise ratios are on the order of 100. IRTF contains mainly solar-metallicity stars with spectral types between F and M, plus some AGB stars and L dwarfs. E-IRTF extends the metalli\-city coverage of IRTF for late-type stars from mainly solar to $-1.7 < \feh < 0.3$. 
Our sample counts 19 stars in common with IRTF 
(in practice 25 XSL spectra) and 
48 stars in common with E-IRTF (61 XSL spectra), 
The cross-match between XSL and IRTF then corresponds to 86 spectra in total,
and 77 when we restrict the sample to XSL spectra with slit-loss corrections.

Figure~\ref{fig:color_irtf} shows the comparison. We find a good agreement with insignificant offsets ($\leqslant 0.007$\,mag) and an initial rms scatter of $0.033$ and $0.029$\,mag, respectively for $(J-H)$ and $(H-K_s)$.
We checked all the outliers individually (respectively 6 and 4 cases for these two colors), by overplotting the IRTF and XSL spectra on one hand, and by inspecting the comparison of the XSL spectra with best-fit stellar models  on the other. One outlier is a Mira that indeed varied in spectral type between the observations. Another is a star listed as abnormal in Table\,\ref{tab:peculiar} (X0878). For all other cases, the IRTF SEDs show signs of a discontinuity between the J and the H band, while the comparison between the XSL and the theoretical spectra essentially excludes errors larger than a few percent. Removing the 6 suspicious cases from the sample, the scatter drops to 0.023\,mag in $(J-H)$ and 0.019\,mag in $(H-K_s)$. Hence a representative value of our internal uncertainties in these near-infrared colors, at least for this subsample, is 2\,\%.


\section{Final data products and availability}
\label{sec:data_prod}

The final data products released as DR2 consist of single-arms spectra,  with linearly sampled wavelengths in the rest-frame.
The spectra are provided as FITS binary tables, each containing three columns: the wavelength (in nm),
the flux spectrum (in erg s$^{-1}$ cm$^{-2}$ \AA$^{-1}$), 
and an error spectrum (in the same units).
A set of keywords in the headers of the FITS files keep track of various steps of the data-reduction and calibration processes, and of results of the quality assessments. A (non-exhaustive) list of these keywords is given in
Table~\ref{table:key_dico}.

\begin{table*}[!ht]
\caption{XSL keywords dictionary (primary header).}
{\centering
\resizebox{\linewidth}{!}{%
\begin{tabular}{l| l l l  c}
\toprule\toprule
General topic & Step 			& Keyword$^{(1)}$ & 	Description  & Default value\\
\toprule
\multirow{3}{2.5cm}{General information} & & OBJECT &  Machine-readable name of the star & \\
& & HNAME &  Human-readable name of the star & \\
& & XSL\_ID & X-shooter Spectral Library unique identifier & \\
& & XSL\_NAME & Internal name for the science spectrum & \\
& & PROV\textit{i} & Originating raw science file(s) & \\
\midrule
\multirow{12}{2.5cm}{Data reduction} & 1D extraction & EXT\_AVG & 	Two 1D-spectra averaged  & F \\
\cmidrule{2-5}
& Response curve	& CAL\_NAME & Internal name for the response curve used & \\
& & CAL\_STAR & Spectro-photometric standard star target & \\
& & CAL\_RAW\textit{i} & Originating raw flux-standard file(s) & \\
\cmidrule{2-5}
& First flux calibration  	& FLUX\_COR & 	First flux calibration applied & F \\
\cmidrule{2-5}
& Flux-loss correction 	& LOSS\_COR & 	Correction for slit-losses applied & F \\
& & STA\_NAME & Internal name for the wide-slit spectrum used & \\
& & STA\_RAW\textit{i} & Originating raw wide-slit file(s) & \\
\midrule
\multirow{15}{2.5cm}{Characterization and quality assurance}  & Rest-frame correction & REST\_COR & Correction to rest-frame applied & F\\
&                      & REST\_VAL & $cz$ value$^{(2)}$ [in \kms]  & \\
\cmidrule{2-5}
&   Barycor values & BARY\_COR & Barycentric radial velocity correction value & \\ 
\cmidrule{2-5}
& Quality flags &  SNR 	& 		Median signal-to-noise ratio & \\
\cmidrule{3-5}
& & WAVY\_UVB &  	Wavy spectrum between 460 and 520 nm & F\\
& & &  	(suspected residuals of the blaze function) & \\
& & WMIN\_UVB & 	Shortest useful wavelength [in nm]  & 300 \\ 
\cmidrule{3-5}
& & HAIR\_VIS & Some narrow spikes in the VIS spectrum & F\\
& & NOIS\_VIS & Noisy VIS spectrum & F \\
& & WAVY\_VIS & Wavy in some part of the VIS spectrum  & F	 \\
& & WMIN\_VIS & 	Shortest useful wavelength [in nm] & 540 \\ 
\cmidrule{3-5}
& & HAIR\_NIR & Some narrow spikes in the NIR spectrum & F \\
& & WAVY\_NIR & Wavy in some part of the NIR spectrum  & F	 \\
\bottomrule
\end{tabular}
}
\label{table:key_dico}
\\
} 
Notes to Table\,\ref{table:key_dico}. (1) The values assigned to the keywords in the headers of the DR2 data
files may depend on the arm (UVB, VIS or NIR), even when this is not explicit in the keyword name. (2) The $cz$ values are barycentric. They are affected by uncertainties directly related to those on the standard wavelength calibration of the Xshooter data (cf. Sect.\,\ref{sec:arm-velo}), as illustrated by differences between arms. 
\end{table*}

The DR2 spectra are made available as two separate sets. The first contains those spectra for which a correction for flux-losses across the spectrograph slit was possible, as described in Section~\ref{sec:flux_loss}). It represents 85\,\% of the observations. We choose to also deliver the remaining 15\% of the spectra, for which the flux-losses could not be estimated and corrected. They are identified via header keyword LOSS\_COR, which is then set to F.
The data can be downloaded from the XSL website, from the CDS interface or from the ESO Phase 3 interface.


\section{Concluding remarks}
\label{sec:concl}

In \citet{Chen14}, we release 246 near-ultraviolet and visible spectra for 237 stars. In this paper, we present the second data release of the X-Shooter Spectral Library, which consists of all the spectra observed as part of this Large Program, in all the three wavelength ranges of the instrument. The 
major addition of this release is the near-infrared data, which extends out to $2.5\,\mu$m. The  simultaneous optical and near-infrared observations in this unique dataset  eliminate a major risk of inconsistencies that has prevailed in previous empirical multi-wavelength libraries based on independent samples in the two wavelength regions.
The library contains 2\,388 spectra of 666 unique stars, at a resolution $R \sim$ 10\,000. This corresponds to 791 UVB spectra, 811 VIS spectra and 786 NIR spectra of these stars.

All the spectra, including those previously released, have been reduced in a homogeneous way with updated algorithms. They were corrected for atmospheric extinction and  telluric absorption, which maximizes the wavelength ranges  suitable for scientific exploitation. In particular, the wings of the water bands of the atmospheres of cool stars in the NIR arm can be measured, and synthetic photometry can be exploited across wavelength regions where it would otherwise be close to meaningless. 

The observational setup and the data reduction were tailored to achieve a good relative flux calibration of the spectra, prioritizing the energy distributions over absolute fluxes. This effort was driven by two main scientific applications foreseen for XSL. In population synthesis calculations, the energy distributions are important in determining the relative fluxes of various stars at different wavelengths. They are also relevant if the data are used to test how well current theoretical stellar models are able to reproduce observed colors and spectral features simultaneously. Flux calibration is notoriously difficult to achieve with slit-fed echelle spectrographs and X-Shooter is no exception. 85\,\% of the DR2 spectra are flux calibrated in the sense that they are corrected for all stable transmission factors of the acquisition chain, and for wavelength-dependent losses due to the narrow widths of the slits that ensured the desired spectral resolution. 
For these 85\,\%, residual flux calibration errors may come from changes in telescope+instrument+sky transmission curves between the observations of the science targets and of the spectrophotometric standard star, and from residual slit losses in broad-slit observations. 
The remaining 15\,\% of the DR2 spectra are not corrected for slit-losss, because the matching broad-slit observations were faulty. 
Systematic differences of the energy distributions with those of previous libraries, if any, are small: the average differences between the synthetic colors measured on XSL spectra (after slit-loss correction) and on spectra from Miles or (E-)IRTF are below 1\,\%; the dispersion in these comparisons indicates typical individual errors on these broad-band colors, outside the main telluric regions, of 2 to 4\,\% in XSL (with some outliers). Anticipating future publications, we note that comparisons with theoretical spectral libraries across the HR-diagram show clear systematic differences that depend on location in that diagram, with excellent matches in some regions and offsets in certain colors in others; this demonstrates that the current flux calibration allows us to address the scientific questions that were planned. We note that it may be possible to improve the spectrophotometry of XSL spectra in the future, at least at optical wavelengths, pending on additional observations with other instruments  \citep[e.g., the integral-field spectrograph MUSE, as suggested by][]{Ivanov2019}. 

The residual errors in the relative flux scales of the spectra of the three arms of X-Shooter are generally smaller than a few percent. However, there are exceptions, and merging the arms for all the DR2 spectra is a non-trivial process. As previously mentioned, simply fitting the observations to (adequately reddened) theoretical spectra does not provide immediately useful re-scaling factors, as available collections of models do not match the observed energy distributions sufficiently well in all parts of the HR diagram (again, the systematics in these trends show that the issues are not artifacts of flux calibration procedures). Work on merging the arm-spectra is ongoing \citep{Verro19}. This work is tied to the development of a spectral interpolator for XSL, which is needed to predict XSL-based spectra along the isochrones of artificial stellar populations. The comparison between interpolated and original library spectra helps identifying and correcting merged spectra for which the initial relative scaling is inadequate.

The DR2 spectra are made available in the restframe, with radial velocity information in the headers. While the restframe calibration is accurate to typically better than 2\,km/s, the radial velocity estimates are affected by larger instrumental instabilities and should be regarded with caution.

Because XSL extends into the near-infrared, it emphasises cool and infrared-luminous stars which otherwise tend to be scarce in stellar libraries. Studying the AGB stars and red supergiants by themselves will help reduce current discrepancies with synthetic spectra based on static and dynamical stellar atmosphere models \citep[e.g.,][]{LanconIAU18}, and will contribute to our understanding of young and intermediate age stellar populations. However, XSL remains a general-purpose library that covers a wide range of stellar types and chemical compositions. Fundamental stellar parameters have already been determined and published \citep{Arentsen19}. We expect XSL will serve a diversity of projects, and in particular improve stellar population synthesis models at a time where many observing facilities concentrate on offering near-infrared instrumentation.


\begin{acknowledgements}
We are grateful to the ESO astronomers who obtained the data presented in this paper in Service Mode operations at Paranal Observatory. This research has made use of the SIMBAD database and the VizieR catalog, both operated at CDS, Strasbourg, France. 
AG is supported by the European Union FP7 programme through ERC grant number 320360. AG is grateful to the IAC where part of this laborious work has been done.
AG and PSB thank the International Space Science Institute in Beijing (ISSI-BJ) for supporting and hosting the meeting of the International Team on “Stellar Libraries of 2020”, during which some discussions contributing to this publication were held. AG and AL thank C\'ecile Loup for useful discussions during this work. AL, PP, AG gratefully acknowledge repeated support by the Programme National de Physique Stellaire (PNPS) of CNRS/INSU, and by the Programme National Cosmology et Galaxies (PNCG) of CNRS/INSU with INP and IN2P3, co-funded by CEA and CNES. 
PC acknowledges support from Fundaç\~{a}o de Amparo à Pesquisa do Estado de S\~{a}o Paulo (FAPESP) through project 2018/05392-8, and Conselho Nacional de Desenvolvimento Científico e Tecnológico (CNPq) through project 310041/2018-0.
JFB and AV acknowledge support from grant AYA2016-77237-C3-1-P from the Spanish Ministry of Economy and Competitiviness (MINECO). 
PSB acknowledges financial support from the Spanish Ministry of Science and Innovation under grant AYA2016-77237-C3-2P. 
AA gratefully acknowledges funding by the Emmy Noether program from the Deutsche Forschungsgemeinschaft (DFG).
We thank the referee for the helpful comments on the text and the website.

\end{acknowledgements}


\bibliographystyle{aa}  
\bibliography{xsl_dr2_biblio}


\begin{appendix}

\section{Input catalogs}
\label{app:input_cats}

Table \ref{tab:inputcats} provides a list of references other than 
Elodie, MILES, NGSL and PASTEL, that were used to select XSL targets. 

\begin{table*}
\begin{center}
\caption[]{Input catalogs for stars not in MILES, NGSL or PASTEL}
\begin{tabular}{lll}
\hline \hline
Object name$^{(1)}$ & Reference & Object category \\
in Tab.\ref{tab:sample} & & \\
\hline
HIP* & \cite{Gray06} & Main sequence stars in the MW \\
HE* & \cite{Yong03} & Very metal poor stars in the MW \\
CD*, CPD* & \cite{Levesque05} & Red supergiants in the MW \\
HD*, HE* & \cite{Bergeat02}, \cite{Christlieb01} & C stars in the MW \\
IRAS* or GCVS names$^{(2)}$ & \cite{Whitelock06}, or GCVS & LPVs in the MW (mostly C) \\
HD\,1638 & \cite{McDonald12}  & K giant in MW (1 object) \\
Miscellaneous & IRTF, \cite{Cushing05}, \cite{Rayner09} & Miscellaneous stars in the MW\\
Misc. or GCVS names$^{(2)}$ & \cite{LW2000} & Red giants, supergiants and LPVs \\
OGLEII\,DIA\,BUL* & \cite{Groen05} & LPVs in Bulge fields \\
BMB* or 2MASSJ* & \cite{BMB84} & LPVs near NGC\,6522 (near the Bulge) \\
CL\,NGC6522\,Arp* or B86* & \cite{Arp65}, \cite{Blanco86} & LPVs near NGC\,6522 (near the Bulge) \\
2MASSJ* & \cite{Zoccali08} & giants toward the Bulge \\
PHS* & \cite{Pompeia08} & metal-poor giants in the LMC (3 objects) \\
SHV* & \cite{HW90} & LPVs in the LMC \\
M2002* & \cite{Massey02} & Red supergiants in the LMC and SMC \\
HV* & \cite{WBF83} & LPVs in the LMC and SMC \\
CL\,NGC121*, CL\,NGC371* & \cite{WBF83} & LPVs in/near SMC clusters \\
ISO\,MCMSJ*  &  \cite{Cioni03} & LPVs in the SMC\\
BBB* & \cite{ReidMould90} & SMC supergiants (2 objects) \\
\hline
\end{tabular}
\label{tab:inputcats}
Notes: (1) This column provides generic names or examples from
the listed references, but in a non-exclusive way. For instance, while
some of the objects with HD* names are from \cite{Bergeat02}, most HD*
objects are from our main reference catalogs (Elodie, MILES, NGSL, PASTEL). (2) By GCVS
names, we refer to the label+constellation used in the General Catalog
of Variable Stars \citep[e.g.,][]{GCVS2017}, such as TU\,Car, V354\,Cen, etc.
\end{center}
\end{table*}

\section{Comments about individual objects or spectra}
\label{app:comments_stars}

In this DR2 of XSL, we have chosen to include spectra of peculiar objects that made it into the collection for a variety of reasons, and some spectra with known observational artifacts, because they may be of interest to some future users. We list the most obvious cases in Table\,\ref{tab:peculiar}. Many of these spectra should not to use in standard population synthesis applications.


\begin{table*}
\caption[]{Peculiar stars, abnormal spectra or deprecated objects}
\label{tab:peculiar}
\centering
\resizebox{1.07\linewidth}{!}{
\begin{tabular}{l| lll}
\toprule\toprule
Category & Spectrum (XSL\_ID) & Star (HNAME) & Comment \\ 
\toprule
\multirow{16}{1.5cm}{Peculiar}  & X0085, X0133 & HD\,96446 & He-rich \\
&X0116 & HD\,57060 & P-Cygni emission in H$\alpha$ \& HeI (1.083$\mu$m); known eclipsing binary \\
&X0190 & OGLEII\,DIA\,BUL-SC19\,2332 & O-rich mira, possible TiO emission near 1.25\,$\mu$m \citep{LanconIAU18} \\
&X0214, X0248 & HD\,190073 & Herbig Ae/Be star, numerous emission lines, \\
  & & & continuum contribution from cool component in the NIR arm \\
&X0357 & CD-69\,1618 & He-rich \\
&X0420 & SV*\,HV\,11223 & S-star signatures confirmed \citep[ZrO, 940\,nm; cf.][]{WBF83} \\
\bottomrule
\end{tabular}
}
\end{table*}


\section{List of XSL stars}
\label{app:list_stars}

Table~\ref{tab:sample} details our list of stars and some of their properties.

\clearpage
\onecolumn
\begin{longtable}[!ht]{lrrccc} 
\kill 
\caption{List of XSL observations (ordered by increasing RA)} \\
\hline \hline
Star      & RA  & DEC  & XSL\_ID & MJD & Arm missing  \\
 & (J2000)  &  (J2000)  &  & & or deprecated  \\ 
 \hline 
 \endfirsthead
 \caption[]{continued} \\
  \hline \hline
 Star      & RA  & DEC  & XSL\_ID & MJD & Arm missing  \\
\hline
 \endhead
%
HD 224926 & 00:01:49.45 & -03:01:39.02 & X0022 & 55112.9969 &  \\
HD 224926 & 00:01:49.45 & -03:01:39.02 & X0224 & 55410.3310 &  \\
HD 224926 & 00:01:49.45 & -03:01:39.02 & X0280 & 55457.2688 &  \\
HD 358 & 00:08:23.26 & +29:05:25.55 & X0025 & 55113.0508 &  \\
HD 1638 & 00:20:28.31 & -30:33:19.88 & X0023 & 55113.0104 &  \\
HD 1638 & 00:20:28.31 & -30:33:19.88 & X0180 & 55407.3051 &  \\
HD 1638 & 00:20:28.31 & -30:33:19.88 & X0258 & 55439.2983 &  \\
HD 1638 & 00:20:28.31 & -30:33:19.88 & X0286 & 55461.3079 &  \\
\end{longtable}


\end{appendix}

\end{document}